\documentclass[acmsmall,screen]{acmart}
\AtBeginDocument{%
  }

\setcopyright{cc}
\setcctype{by}
\acmDOI{10.1145/3808144}
\acmYear{2026}
\acmJournal{PACMSE}
\acmVolume{3}
\acmNumber{FSE}
\acmArticle{FSE137}
\acmMonth{7}
\acmSubmissionID{fse26mainb-p535-p}
\received{2026-02-25}
\received[accepted]{2026-03-24}

\usepackage{algorithmic}
\usepackage{xcolor}
\usepackage{enumitem}
\usepackage{booktabs}
\usepackage{tabularx}
\usepackage{multirow}
\usepackage{array}
\usepackage{graphicx}
\usepackage{subcaption}
\usepackage{adjustbox}
\usepackage{makecell}
\usepackage[most]{tcolorbox}
\usepackage[switch]{lineno} % 引入行号包
\usepackage{url}
\usepackage{textcomp}
\usepackage[normalem]{ulem} % 导入 ulem 包，normalem 选项可以防止它影响 \emph 命令

\newcommand{\add}[1]{{#1}} % 新增内容使用红色
\newcommand{\note}[1]{} % 新增内容使用红色
\newcommand{\swrbench}{\textit{SWR-Bench}}

\newcommand{\changeprs}{\textit{Change-PRs}}
\newcommand{\cleanprs}{\textit{Clean-PRs}}
\newcommand{\cleanpr}{\textit{Clean-PR}}
\newcommand{\cas}{\textit{change-actions}}
\newcommand{\ca}{\textit{change-action}}
\newcommand{\CAS}{\textit{Change-Actions}}

\newcommand{\codein}[1]{\texttt{#1}}

\newcolumntype{C}[1]{>{\RaggedRight\arraybackslash}p{#1}}
\NewDocumentCommand{\mybox}{ m m }{
    \begin{tcolorbox}[enhanced, frame hidden, boxsep=0pt]
    \emph{\textbf{#1}: #2}
    \end{tcolorbox}
}

\begin{document}

% \title{SWR-Bench: Assessing LLM Performance in Real-World Code Review Comment Generation}

% \author{Zhengran Zeng}
% \authornote{Both authors contributed equally to this research.}
% \email{zhengranzeng@stu.pku.edu.cn}
% \orcid{0009-0009-8422-4522}
% \author{Ruikai Shi}
% \authornotemark[1]
% \email{rkshi25@stu.pku.edu.cn}
% \orcid{0009-0001-2028-670X}
% \affiliation{%
%   \institution{Peking University}
%   \state{Beijing}
%   \country{China}
% }
% \author{Keke Han}
% \orcid{0009-0003-1843-8545}
% \email{kkhan25@stu.pku.edu.cn}
% \author{Yixin Li}
% \orcid{0009-0005-0492-2234}
% \email{leason_lyx@stu.pku.edu.cn}
% \affiliation{%
%   \institution{Peking University}
%   \state{Beijing}
%   \country{China}
% }
% \author{Kaicheng Sun}
% \orcid{0009-0007-0856-9049}
% \email{sunkaicheng@mail.nwpu.edu.cn}
% \affiliation{%
%   \institution{Northwestern Polytechnical University}
%   \state{Xian}
%   \country{China}
% }
% \author{Yidong Wang}
% \orcid{0009-0007-9969-8259}
% \email{2301110730@stu.pku.edu.cn}
% \author{Zhuohao Yu}
% \orcid{0009-0000-8256-8588}
% \email{zyu@stu.pku.edu.cn}
% \affiliation{%
%   \institution{Peking University}
%   \state{Beijing}
%   \country{China}
% }
% \author{Rui Xie}
% \orcid{0000-0002-1756-7746}
% \authornote{Those authors are the corresponding authors.}
% \email{ruixie@pku.edu.cn}
% \author{Wei Ye}
% \orcid{0000-0002-9331-4716}
% \authornotemark[2]
% \email{wye@pku.edu.cn}
% \author{Shikun Zhang}
% \orcid{0000-0002-8576-2674}
% \authornotemark[2]
% \email{zhangsk@pku.edu.cn}
% \affiliation{%
%   \institution{Peking University}
%   \state{Beijing}
%   \country{China}
% }

\title{SWR-Bench: Assessing LLM Performance in Real-World Code Review Comment Generation}

\author{Zhengran Zeng}
\authornote{Both authors contributed equally to this research.}
\orcid{0009-0009-8422-4522}
\affiliation{%
  \institution{Peking University}
  \city{Beijing}
  \country{China}
}
\email{zhengranzeng@stu.pku.edu.cn}

\author{Ruikai Shi}
\authornotemark[1]
\orcid{0009-0001-2028-670X}
\affiliation{%
  \institution{Peking University}
  \city{Beijing}
  \country{China}
}
\email{rkshi25@stu.pku.edu.cn}

\author{Keke Han}
\orcid{0009-0003-1843-8545}
\affiliation{%
  \institution{Peking University}
  \city{Beijing}
  \country{China}
}
\email{kkhan25@stu.pku.edu.cn}

\author{Yixin Li}
\orcid{0009-0005-0492-2234}
\affiliation{%
  \institution{Peking University}
  \city{Beijing}
  \country{China}
}
\email{leason_lyx@stu.pku.edu.cn}

\author{Kaicheng Sun}
\orcid{0009-0007-0856-9049}
\affiliation{%
  \institution{Northwestern Polytechnical University}
  \city{Xian}
  \country{China}
}
\email{sunkaicheng@mail.nwpu.edu.cn}

\author{Yidong Wang}
\orcid{0009-0007-9969-8259}
\affiliation{%
  \institution{Peking University}
  \city{Beijing}
  \country{China}
}
\email{2301110730@stu.pku.edu.cn}

\author{Zhuohao Yu}
\orcid{0009-0000-8256-8588}
\affiliation{%
  \institution{Peking University}
  \city{Beijing}
  \country{China}
}
\email{zyu@stu.pku.edu.cn}

\author{Rui Xie}
\authornote{Those authors are the corresponding authors.}
\orcid{0000-0002-1756-7746}
\affiliation{%
  \institution{Peking University}
  \city{Beijing}
  \country{China}
}
\email{ruixie@pku.edu.cn}

\author{Wei Ye}
\authornotemark[2]
\orcid{0000-0002-9331-4716}
\affiliation{%
  \institution{Peking University}
  \city{Beijing}
  \country{China}
}
\email{wye@pku.edu.cn}

\author{Shikun Zhang}
\authornotemark[2]
\orcid{0000-0002-8576-2674}
\affiliation{%
  \institution{Peking University}
  \city{Beijing}
  \country{China}
}
\email{zhangsk@pku.edu.cn}

% Zhengran Zeng (Peking University) <>
% Ruikai Shi (Peking University) <rkshi25@stu.pku.edu.cn>
% Keke Han (Peking University) <kkhan25@stu.pku.edu.cn>
% Yixin Li (Peking University) <leason_lyx@stu.pku.edu.cn>
% Kaicheng Sun (Northwestern Polytechnical University) <sunkaicheng@mail.nwpu.edu.cn>
% Yidong Wang (Peking University) <2301110730@stu.pku.edu.cn>
% Zhuohao Yu (Peking University) <zyu@stu.pku.edu.cn>
% Rui Xie (Peking University) <ruixie@pku.edu.cn>
% Wei Ye (Peking University) <wye@pku.edu.cn>
% Shikun Zhang (Peking University) <zhangsk@pku.edu.cn>

% \author{
%     % 第一行作者
%     \IEEEauthorblockN{
%         Zhengran Zeng\textsuperscript{1}*,
%         Ruikai Shi\textsuperscript{1}*,
%         Keke Han\textsuperscript{1},
%         Yixin Li\textsuperscript{1},
%         Kaicheng Sun\textsuperscript{2}
%     }
%     % 第二行作者 (如果需要)
%     \IEEEauthorblockN{
%         Yidong Wang\textsuperscript{1},
%         Zhuohao Yu\textsuperscript{1},
%         Rui Xie\textsuperscript{1},
%         Wei Ye\textsuperscript{1},
%         Shikun Zhang\textsuperscript{1}
%     }
    
%     % \vspace{1.2em} % 可选：微调作者和机构之间的垂直距离
%     % 机构信息
%     \IEEEauthorblockA{
%         \textsuperscript{1}Peking University, Beijing, China
%     }
%     \IEEEauthorblockA{
%         \textsuperscript{2}Northwestern Polytechnical University, Xian, China
%     }
%     \thanks{\textit{*These authors contributed equally to this work.}}
% }

% \renewcommand{\shortauthors}{Trovato et al.}
\renewcommand{\shortauthors}{Zhengran Zeng, Ruikai Shi, Keke Han, Yixin Li, KC Sun, YD Wang, ZH Yu, Rui Xie, Wei Ye and Shikun Zhang}
\begin{CCSXML}
<ccs2012>
   <concept>
       <concept_id>10011007.10011074.10011111.10011696</concept_id>
       <concept_desc>Software and its engineering~Maintaining software</concept_desc>
       <concept_significance>500</concept_significance>
       </concept>
 </ccs2012>
\end{CCSXML}

\ccsdesc[500]{Software and its engineering~Maintaining software}

\begin{abstract}
Automated Code Review (ACR) is crucial for software quality, yet existing benchmarks often fail to reflect real-world complexities, hindering the evaluation of modern Large Language Models (LLMs). Current benchmarks frequently focus on fine-grained code units, lack complete project context, and use inadequate evaluation metrics. To address these limitations, we introduce \swrbench{}, a new benchmark comprising 1000 manually verified Pull Requests (PRs) from GitHub, offering PR-centric review with full project context. \swrbench{} employs an objective LLM-based evaluation method that aligns strongly with human judgment ($\sim$90\% agreement) by verifying if issues from a structured ground truth are covered in generated reviews. Our systematic evaluation of mainstream ACR tools and LLMs on \swrbench{} reveals that current systems underperform, and ACR tools are more adept at detecting functional errors. Subsequently, we propose and validate a simple multi-review aggregation strategy that significantly boosts ACR performance, increasing F1 scores by up to 43.67\%. Our contributions include the \swrbench{} benchmark, its objective evaluation method, a comprehensive study of current ACR capabilities, and an effective enhancement approach, offering valuable insights for advancing ACR research.
\end{abstract}

\keywords{Automated Code Review, Large Language Models, Benchmark}

\maketitle

\section{Introduction}
\label{sec:introduction}

Code Review (CR) is an indispensable quality assurance practice in software development~\cite{codereview_survey,codereview_survey_2,codereview,codereview_2,codeagentsurvey}, aimed at identifying and rectifying potential issues in code changes before integration. While crucial for enhancing code quality, traditional manual code review faces significant hurdles in modern software development. The increasing scale and complexity of projects mean that manual reviews are time-consuming, contributing to development costs and potential delays in feature releases~\cite{codereview_challenge,codereview_challenge_2,codereview_challenge_3}. 
To mitigate these challenges, Automated Code Review (ACR) Comment Generation has emerged as a vital field and a prominent research area within software engineering~\cite{acr,Trans-ReviewData,AutoTransformData,T5-ReviewData,codereviewer}, driving the creation of numerous tools and datasets. These ACR works strive to improve the efficiency and effectiveness of the code review process by shortening feedback cycles, reducing the manual burden on developers, and improving the consistency and scope of reviews, thus playing a crucial role in streamlining software development workflows and maintaining high standards of code quality.
More recently, the explosive advancements in Large Language Models (LLMs) have catalyzed a significant shift in ACR research~\cite{BitsAI-CR,hybird_review,acr_llm_3,acr_llm_4}, with a growing emphasis on LLM-based methodologies. 

\begin{figure}[htbp] % 'h'ere, 't'op, 'b'ottom, 'p'age of floats. [!htbp] is more forceful.
    \centering % 中心对齐整个 figure 环境
    \begin{subfigure}[b]{0.48\linewidth} % 子图1，宽度为文本宽度的48%
        \centering
        \includegraphics[width=0.9\linewidth]{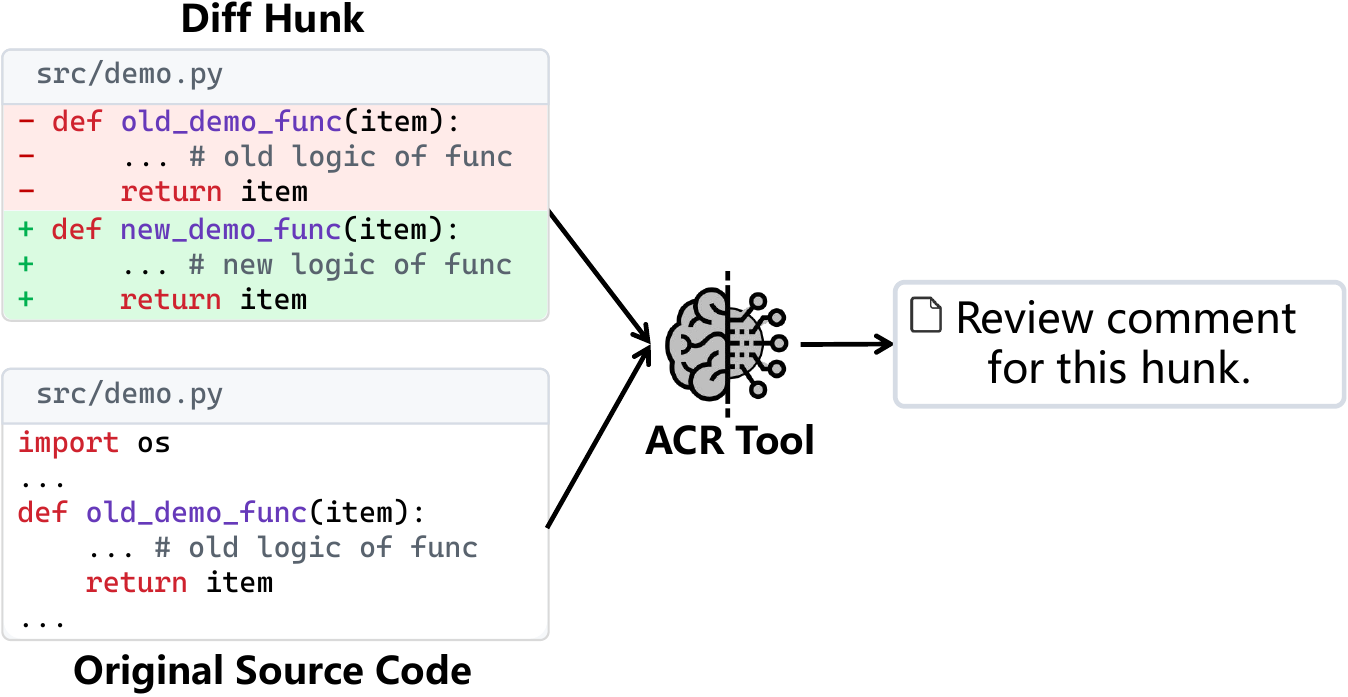} % 替换为你的第一个图片文件名
        \caption{Traditional hunk-level review with limited context.}
        \label{fig:hunk_paradigm}
    \end{subfigure}
    \hfill % 在两个子图之间添加水平弹性空间，使它们分开
    \begin{subfigure}[b]{0.48\linewidth} % 子图2，宽度为文本宽度的48%
        \centering
        \includegraphics[width=0.9\linewidth]{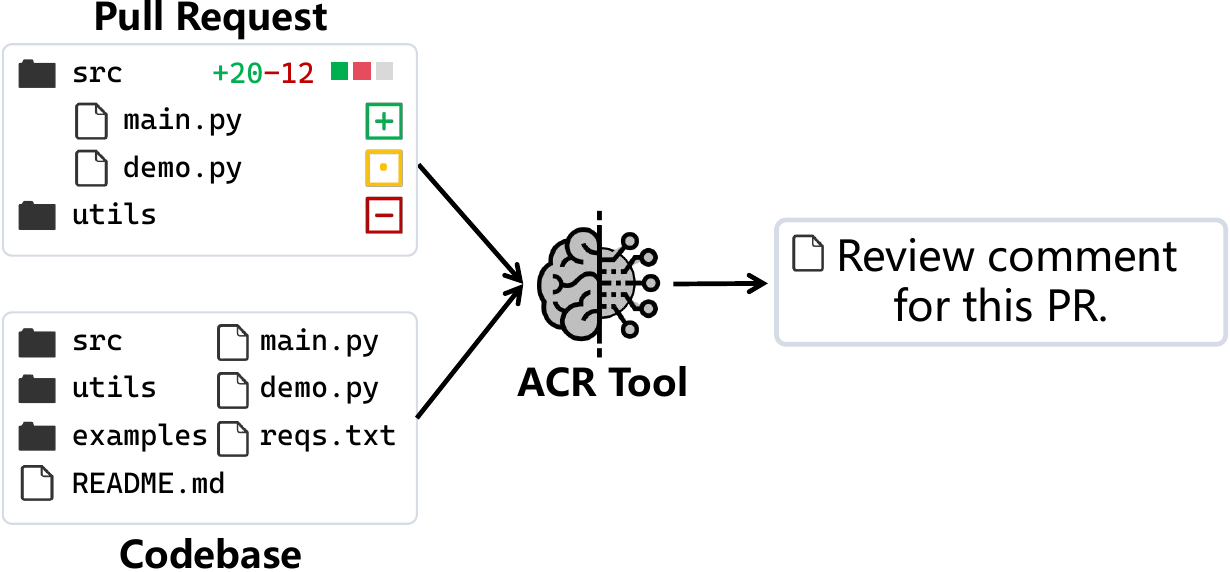} % 替换为你的第二个图片文件名
        \caption{SWR-Bench PR-level review with full codebase context.}
        \Description{}
        \label{fig:pr_paradigm}
    \end{subfigure}
    \caption{A comparison of review paradigms.} 
    \label{fig:review_paradigms} % 整个大图的标签（可选，如果需要引用整个组合图）
\end{figure}

However, a substantial portion of existing ACR benchmarks~\cite{Trans-ReviewData,AutoTransformData,T5-ReviewData,codereviewer} were established primarily for deep learning models that predate the widespread capabilities and sophisticated understanding of modern LLMs. This temporal and methodological gap presents challenges in evaluating the true potential of contemporary LLM-based ACR tools using these benchmarks. This discrepancy is particularly evident in the fundamental setup of the review task itself, as illustrated in Figure~\ref{fig:review_paradigms}.
Specifically, existing representative benchmarks~\cite{Trans-ReviewData,T5-ReviewData,AutoTransformData,codereviewer} primarily center their evaluation on isolated code changes, or diff hunks, with limited contextual information, such as only a few related source code files (see Figure~\ref{fig:hunk_paradigm}). This hunk-based approach fails to mirror real-world developer practices, where an entire Pull Request (PR) is reviewed as a cohesive unit. 
Consequently, this narrow scope can lead to missing critical inter-dependency bugs.% (as we will demonstrate in Section~\ref{sec:background}). 
Meanwhile, these benchmarks typically do not provide the complete project codebase as necessary context, making it difficult for ACR tools to understand the global impact of code changes. In contrast, our paradigm provides the full context (Figure~\ref{fig:pr_paradigm}).
More critically, the evaluation methodologies for these benchmarks are also problematic. They generally rely on traditional natural language generation metrics such as $Exact\text{-}Match$~\cite{Trans-ReviewData}, $Bleu$ scores~\cite{BLEU} and simple LLM-based ratings~\cite{LLM-as-judge_3,pandalm,hybird_review}. However, metrics like $Bleu$ primarily measure textual similarity, while the reliability of LLM ratings is often questioned due to potential inherent biases; both approaches have been shown to be severely inadequate in assessing whether the crucial issues are truly identified in code review comments~\cite{metric_weakness,metric_weakness_2}. Furthermore, some studies~\cite{BitsAI-CR,acr} utilize expensive human annotation, but this approach is difficult to scale. 
% Consequently, existing ACR benchmarks commonly suffer from three key limitations: a narrow, hunk-level review scope detached from real-world PR-centric practices; insufficient project context for assessing the global impact of changes; and inadequate evaluation metrics that fail to capture semantic accuracy.

Consequently, evaluations on current benchmarks often misrepresent the true capabilities and practical value of LLMs in complex, real-world code review. To bridge this critical gap, we introduce \swrbench{} (\textbf{S}oft\textbf{w}are \textbf{R}eview \textbf{Bench}mark), a software review benchmark designed for more realistic evaluation. Specifically, \swrbench{} comprises 1000 manually verified PR instances from GitHub open-source projects. It directly confronts the issues in existing benchmarks through its core design features:

% \add{Consequently, evaluations on current benchmarks often misrepresent the true capabilities and practical value of LLMs in complex, real-world code review. To bridge this critical gap, we argue that advancing the entire end-to-end ACR pipeline, which includes downstream tasks like automated code refinement, must begin with perfecting the foundational stage: \textbf{review comment generation}. An effective ACR tool must first generate precise comments that identify the correct location, specify the error, and suggest a solution. This high-quality output then serves as the direct and necessary input for any subsequent automated refinement. Therefore, our work focuses on rigorously evaluating this crucial first step. We introduce \swrbench{} (\textbf{S}oft\textbf{w}are \textbf{R}eview \textbf{Bench}mark), a benchmark designed to assess this foundational capability in a realistic setting. Specifically, \swrbench{} comprises 1000 manually verified PR instances from GitHub open source projects. It directly confronts the issues in existing benchmarks through its core design features:}

\begin{itemize}[leftmargin=*, nosep]
    \item \textbf{PR-centric Review:}
    Each code review instance is based on a complete PR. This setup is intentionally more challenging, yet more aligned with real-world practices, as we aim to assess the ability of ACR tools for end-to-end code review. This includes identifying areas requiring review, pinpointing specific issues, and generating meaningful review comments.
    \item \textbf{Comprehensive Context:}
    To simulate a realistic review environment, our benchmark provides all relevant code changes (commits) and a snapshot of the entire project repository for each PR, offering comprehensive context that enables tools to accurately assess the global impact of modifications.
    \item \textbf{Objective LLM-based Evaluation:}
    We employ an automated evaluation where an LLM objectively verifies if issues from a structured ground truth report are covered in the generated report. This method offers a more objective and semantically relevant assessment than subjective ``LLM-as-judge'' ratings~\cite{LLM-as-judge_3,pandalm,hybird_review} or text-similarity metrics~\cite{Trans-ReviewData,BLEU}.
\end{itemize}

After establishing \swrbench{}, we first validated its evaluation method, confirming strong ($\sim$90\%) agreement with human judgment. We then systematically evaluated mainstream ACR tools and leading LLMs on \swrbench{}, uncovering key performance aspects in realistic scenarios: 1) current tools and LLMs do not yet perform sufficiently well; 2) furthermore, we found that LLMs trained with a reasoning-focused approach exhibit better code review capabilities; and finally, 3) existing tools are more adept at detecting functional errors, such as bugs, compared to identifying non-functional issues like outdated documentation. Motivated by these observations, we further introduce a simple enhancement scheme for LLM-based automated code review. This straightforward strategy empowers an LLM to synthesize feedback from multiple review sources by analyzing, filtering valid points, and discarding erroneous suggestions to produce a superior final report. Experiments demonstrate this strategy substantially boosts ACR performance (increasing issue detection $F1$ score by up to 43.67\%), suggesting promising avenues for future research. Therefore, our main contributions in this work include:

\begin{enumerate}[label=\arabic*., wide, labelindent=0pt, leftmargin=*]
    \item \textbf{Benchmark:} We introduce \swrbench{}, a benchmark comprising 1,000 real-world PRs with full project context, providing a more realistic and challenging evaluation platform for ACR systems.
    \item \textbf{Evaluation Method:} We developed and validated an objective LLM-based evaluation method to objectively assess ACR quality by comparing tool outputs against human ground truth.
    \item \textbf{Study:} We systematically assessed mainstream LLMs and ACR tools on \swrbench{}, analyzing their performance, strengths, and limitations in practical code review scenarios.
    \item \textbf{Approach:} We developed and validated a simple multiple review strategy that significantly improves the code review performance of ACR tools on \swrbench{}.
\end{enumerate}
% Ultimately, this work offers a more realistic platform to advance automated code review. It aims to align research with real-world needs, helping to identify current bottlenecks and develop ACR tools that are truly effective in practical development.

\section{Background and Related Work}
\label{sec:background}

\subsection{Modern Code Review}
\label{sec:mcr}

\begin{figure}[htbp] % htbp 是图片位置参数建议：here, top, bottom, page
    \centering
    % 请确保您的LaTeX编译环境中有名为 review-process.png, review-process.jpg 或 review-process.pdf 的图片文件
    % 您可能需要调整 width 参数以适应您的文档布局
    \includegraphics[width=0.50\linewidth]{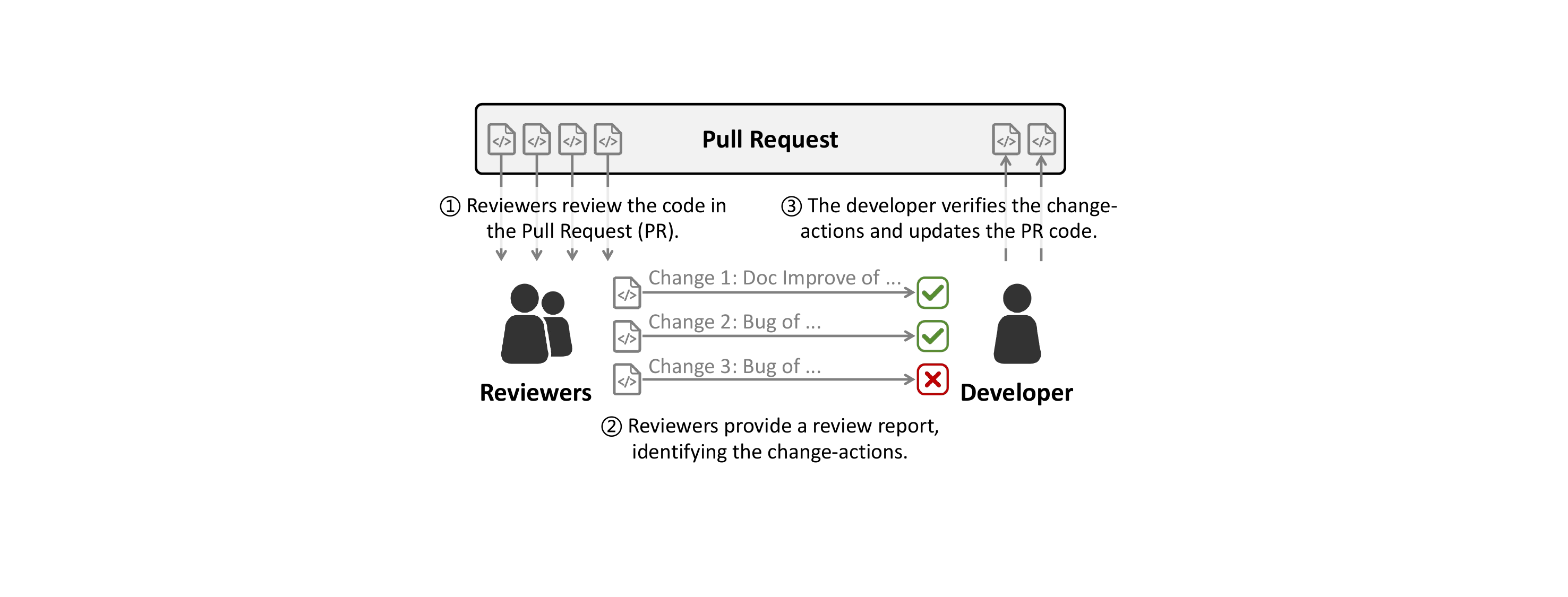}
    \caption{A simplified code review process.}
    \Description{}
    \label{fig:review-process}
\end{figure}

Modern Code Review (MCR) is a widely adopted quality assurance practice in contemporary software development~\cite{codereview_survey,codereview_survey_2}, crucial for enhancing code quality, fostering knowledge sharing, and ensuring adherence to project standards. It typically involves developers submitting code changes, often through PRs, for scrutiny by their peers before integration into the main codebase.

A simplified, yet representative, MCR process is depicted in Figure~\ref{fig:review-process}. This process generally unfolds in three key stages: 1) reviewers scrutinize the code submitted in the pull request; 2) following their examination, reviewers provide a review report, identifying specific areas requiring modification, which we term ``\cas{}''; 3) the developer then verifies these \textit{change-}\textit{actions} and updates the PR code accordingly. The fundamental unit of actionable feedback within this process is what we term a \ca{}. We define an effective \ca{} as a complete directive that specifies the location(s) of the issue, describes the underlying problem, and proposes a tangible solution. Therefore, a single \ca{} precisely reflects a specific issue and its resolution. An ACR tool capable of generating correct \cas{} can thus effectively guide developers in improving code quality. It is also important to note that a \ca{} is not strictly tied to a specific line of code; it can be conceptual and address a high-level concern spanning multiple code locations, such as an architectural flaw. We use the term `change` rather than `fix` or `defect` because reviewer intent often extends beyond rectifying bugs to include non-functional improvements in readability, maintainability, or design, which are not traditional defects but are vital for software quality~\cite{pr_change_taxonomy_14,pr_change_taxonomy}.

The nature of these \cas{} varies significantly, reflecting diverse reviewer considerations. To classify them, we adopt the detailed taxonomy from established literature shown in Table~\ref{tab:change_types_condensed}~\cite{pr_change_taxonomy_14,pr_change_taxonomy}. We selected this taxonomy because it is a reliable standard that has been consistently used across multiple studies in the Automated Code Review (ACR) domain. It distinguishes between two high-level categories: \textbf{Evolutionary} changes (improving future maintainability) and \textbf{Functional} changes (altering software behavior), which are further decomposed into 11 fine-grained types. Understanding this classification is fundamental for analyzing the effectiveness of both manual and automated review processes.

Furthermore, it is important to note that in this work, we focus specifically on the task of review comment generation, which we consider a critical prerequisite for end-to-end code review automation. The quality of generated comments directly determines the success of downstream tasks like automated code refinement, as precise feedback is the necessary input for any subsequent modifications.

\begin{table}[htbp]
\centering
\caption{The taxonomy of \cas{} under code review, adapted from~\cite{pr_change_taxonomy_14,pr_change_taxonomy}. \note{AQ5: }\add{Note: All \cas{} are classified into leaf-node categories only.}}
\label{tab:change_types_condensed}
\footnotesize
\begin{adjustbox}{width=0.90\linewidth}
\begin{tabular}{|l|p{0.88\linewidth}|}
% \toprule
\hline
\textbf{ID} & \textbf{Type \& Description } \\
% \midrule
\hline\hline
% ---Evolutionary ---
\textbf{E}     & \textbf{Evolutionary:} Improving code's future maintainability and structure, not its function.\\
\textbf{E.1}   & \textbf{Documentation:} Modifying in-code information for better human understanding. \\
\textbf{E.1.1} & \textbf{Textual:} Changing textual code elements \add{such as inline comments and variable/function names}. \\
\textbf{E.1.2}& \textbf{Language Supported:} Using language features for documentation purposes. \\
\textbf{E.2}   & \textbf{Visual Representation:} Modifying code layout for better style and readability. \\ % 主类别信息省略
\textbf{E.3}   & \textbf{Structure:} Altering the project's organization or architecture. \\ % 主类别信息省略
\textbf{E.3.1} & \textbf{Organization:} Reorganizing code by moving or removing parts. \\
\textbf{E.3.2} & \textbf{Solution Approach:} Altering implementation methods or adding maintainability elements (e.g., tests). \\
% \midrule
\hline\hline
% --- Functional ---
\textbf{F}   & \textbf{Functional:} Altering the software's behavior or interactions. \\

\textbf{F.1}   & \textbf{Interface:} Modifying interactions between codebase parts. \\
\textbf{F.2}   & \textbf{Logic:} Changing the code's logical operations or algorithms. \\ % 主类别信息省略
\textbf{F.3}   & \textbf{Resource:} Altering how variables or resources are managed. \\ % 主类别信息省略
\textbf{F.4}   & \textbf{Check:} Modifying checks for unhandled states/conditions. \\ % 主类别信息省略
\textbf{F.5}   & \textbf{Support:} Adjusting interactions with external support systems or libraries. \\ % 主类别信息省略
\textbf{F.6}   & \textbf{Larger Defects:} \note{AQ5: }\add{PRs that were ultimately \textit{not merged} (i.e., rejected suggestions), typically involving major, extensive functional issues or missing features that required fundamental rethinking.} \\ % 主类别信息省略
% \bottomrule
\hline
\end{tabular}
\end{adjustbox}
\end{table}

\subsection{Automatic Code Review}
\label{subsec:acr}
While modern code review, as described above, is a vital software quality practice, manual code review faces efficiency and consistency challenges~\cite{codereview_challenge,codereview_challenge_2,codereview_challenge_3}, motivating the development of ACR, aiming to use tools to assist or replace parts of manual review, thereby improving efficiency, shortening feedback cycles, and enhancing consistency~\cite{acr}.

Specifically, ACR technology has progressed through several stages. Initially, development focused on rule-based and static analysis tools, such as PMD~\cite{PMD} and SonarQube~\cite{SonarQube}, which are effective but can be rigid and prone to false positives. Subsequently, machine learning and deep learning approaches~\cite{Trans-ReviewData,T5-ReviewData,AutoTransformData,codereviewer} emerged for tasks like code change quality assessment~\cite{CodeAgent,hybird_review,PRReview,deepjit,coderujb}, comment generation~\cite{codereviewer}, and automated code repair~\cite{defects4j,swrbench,agentless}; while these methods can learn complex patterns, they may still struggle with deep code intent and complex contexts. More recently, large language model and AI agent-based methods~\cite{BitsAI-CR,hybird_review,acr_llm_3,acr_llm_4,LLM_codereview,CodeAgent}, have demonstrated strong code understanding and generation capabilities, enabling more natural comments and complex reasoning. Cutting-edge work in this area, such as CodeAgent~\cite{CodeAgent}, explores multi-agent systems for collaborative review, while commercial tools like PR-Review~\cite{PRReview} focus on providing targeted, actionable feedback that dynamically learns team norms, aiming to address new challenges in AI-assisted coding like feedback redundancy and unclear prioritization.
Despite the potential of ACR, especially LLM/Agent-based methods, their real-world effectiveness in improving review quality and efficiency remains a key question. This highlights an urgent need for more effective and reliable evaluation methods and benchmarks, as current assessment approaches~\cite{Trans-ReviewData,AutoTransformData,T5-ReviewData,codereviewer,hybird_review} may not adequately capture the performance of these advanced models in complex, realistic code review scenarios.

\subsection{Code Review Benchmarks}
\label{subsec:cr_benchmarks}

To evaluate various ACR techniques, the research community has constructed several specialized code review benchmarks. Table~\ref{tab:benchmark_comparison} provides a comparative overview of representative benchmarks, assessing them based on their data source, scale, fundamental review unit, the scope of provided context, and evaluation methodology. This comparison highlights a significant gap in existing resources, which our work, \swrbench{}, aims to fill.

% Notable among these are a series of datasets by Tufano et al.~\cite{Trans-ReviewData,AutoTransformData,T5-ReviewData} and collaborators (including \textit{Trans-ReviewData}~\cite{Trans-ReviewData}, \textit{AutoTransformData}~\cite{AutoTransformData}, \textit{T5-ReviewData}~\cite{T5-ReviewData}), and the \textit{CodeReviewer}~\cite{codereviewer} dataset by Li et al.

\begin{table}[htbp] % Use [htbp] for better float placement
\centering
\caption{Comparison of Representative Code Review Benchmarks.}
\label{tab:benchmark_comparison}
\resizebox{0.8\textwidth}{!}{
\begin{tabular}{|lccccc|}
% \toprule % Professional top rule
\hline
\textbf{Dataset} & \textbf{Source} & \textbf{Size} & \textbf{Unit} & \textbf{Context} & \textbf{Evaluation} \\
% \midrule % Rule between header and data
\hline\hline
\textit{Trans-Review-Data}~\cite{Trans-ReviewData} & Github, Gerrit & 1,719 & Method & None & Bleu \\
\textit{AutoTransform-Data}~\cite{AutoTransformData} & Gerrit & 14,750 & Method & None & Bleu \\
\textit{T5-Review-Data}~\cite{T5-ReviewData} & Github, Gerrit & 17,194 & Method & None & Bleu \\
\textit{Code-Reviewer-Data}~\cite{codereviewer} & Github & 10,000 & Diff Hunk & None & Bleu, Human Eval. \\
\textit{CR-Agent-Data}~\cite{CodeAgent} & Github & 3,545 & Commit & Related Source Code & Human Eval. \\
\textit{Hybrid-Review-Data}~\cite{hybird_review} & Github & 1,245 & Diff Hunk & Related Source Code & LLM Scoring \\
% \midrule % Rule to separate our work
\hline
\textbf{\swrbench{}} & \textbf{Github} & \textbf{1,000} & \textbf{Pull Request} & \textbf{Complete Codebase} & \textbf{Objective LLM Eval.} \\
% \bottomrule % Professional bottom rule
\hline
\end{tabular}
}
\end{table}

\textit{Tufano et al.'s datasets}~\cite{Trans-ReviewData,AutoTransformData,T5-ReviewData}, typically sourced from Gerrit and GitHub projects, center on method-level triplets (i.e., the submitted method, reviewer comments, and the modified method). These are used to evaluate tasks like method-level code transformation, implementing changes based on comments, and generating comments for methods. 
% Meanwhile, those benchmarks primarily use $Exact\text{-}Match$, $Bleu$ and $CodeBleu$~\cite{codebleu} metrics for evaluation.
As indicated in Table~\ref{tab:benchmark_comparison}, these benchmarks provide no external code context and primarily rely on text-similarity metrics like $Bleu$ for evaluation.

% \textbf{The CodeReviewer dataset}~\cite{codereviewer} is also from GitHub pull requests, and its core evaluation unit is single diff hunk (i.e., contiguous modified lines). \textit{CodeReviewer} supports tasks such as code change quality estimation (classifying if a diff hunk has defects), review comment generation for diff hunks, and code improvement based on comments, with metrics including classification $Accuracy$, $F1$ score, $Bleu$, and $Exact\text{-}Match$, while some later studies~\cite{hybird_review} also report performance using simple LLM-based ratings.

\textit{Code-Reviewer-Data}~\cite{codereviewer} shifts the focus to a single diff hunk (i.e., a contiguous block of modified lines). While closer to a review task, it still operates on fragments of a change rather than the whole. Its evaluation metrics also include BLEU and Exact-Match, with some later studies~\cite{hybird_review} adopting simple LLM-based scoring. More recent benchmarks like \textit{CR-Agent-Data}~\cite{CodeAgent} and \textit{Hybrid-Review-Data}~\cite{hybird_review} begin to incorporate related source code as context, but still fall short of providing the complete project environment.

While valuable, these benchmarks have limitations compared to real-world code review. As summarized in Table~\ref{tab:benchmark_comparison}, they consistently lack a PR-level scope and comprehensive context, instead focusing on isolated methods or diff hunks. This hinders evaluation of an ACR tool's ability to manage real-world complexity. Furthermore, these benchmarks predominantly rely on inadequate evaluation metrics, such as $Exact\text{-}Match$ and $Bleu$ or simple LLM-based ratings metric, which are known to correlate poorly with human judgment~\cite{passk,mtbench}, thus leading to potentially misleading assessments of tool efficacy.

Beyond these public benchmarks, some studies~\cite{BitsAI-CR,acr} leverage proprietary internal data for ACR research, validating tools through manual human assessment. While such efforts provide valuable, deep insights, the private nature of this data and the prohibitive cost of manual validation make these approaches unscalable and difficult for the broader community to build upon. This leaves a clear void for a publicly available, context-rich, and scalably evaluable benchmark.

% Consequently, existing public benchmarks, despite their merits, often have significant limitations in scope and evaluation methodology. Evaluations based on them are unlikely to accurately gauge the practical performance of ACR tools, especially sophisticated LLM/Agent models. To this end, we introduces a benchmark focused on complete pull requests, incorporating full project context, and utilizing an objective LLM-as-judge evaluation against ground truth. This initiative seeks to provide the community with a more robust and relevant evaluation platform, thereby fostering the development of more potent ACR tools.

In summary, as Table~\ref{tab:benchmark_comparison} illustrates, existing public benchmarks consistently lack PR-level scope, comprehensive project context, and reliable evaluation metrics, three deficiencies that limit their ability to accurately assess modern LLM-based ACR tools. Meanwhile, proprietary benchmarks with manual evaluation, though insightful, are neither scalable nor reproducible. These gaps motivate the design of \swrbench{}, which we detail in the following section.

\section{SWR-Bench}

This section details our methodology for constructing the automatic code review benchmark. % The process includes selecting and filtering data sources, leveraging LLMs for preliminary annotation, and ensuring the quality and representativeness of the final dataset through rigorous manual verification.

\subsection{Benchmark Construction}
\label{subsec:benchmark_construction}

\begin{figure}[htbp] % htbp 是图片位置参数建议：here, top, bottom, page
\centering
% 请确保您的LaTeX编译环境中有名为 review-process.png, review-process.jpg 或 review-process.pdf 的图片文件
% 您可能需要调整 width 参数以适应您的文档布局
\includegraphics[width=0.65\linewidth]{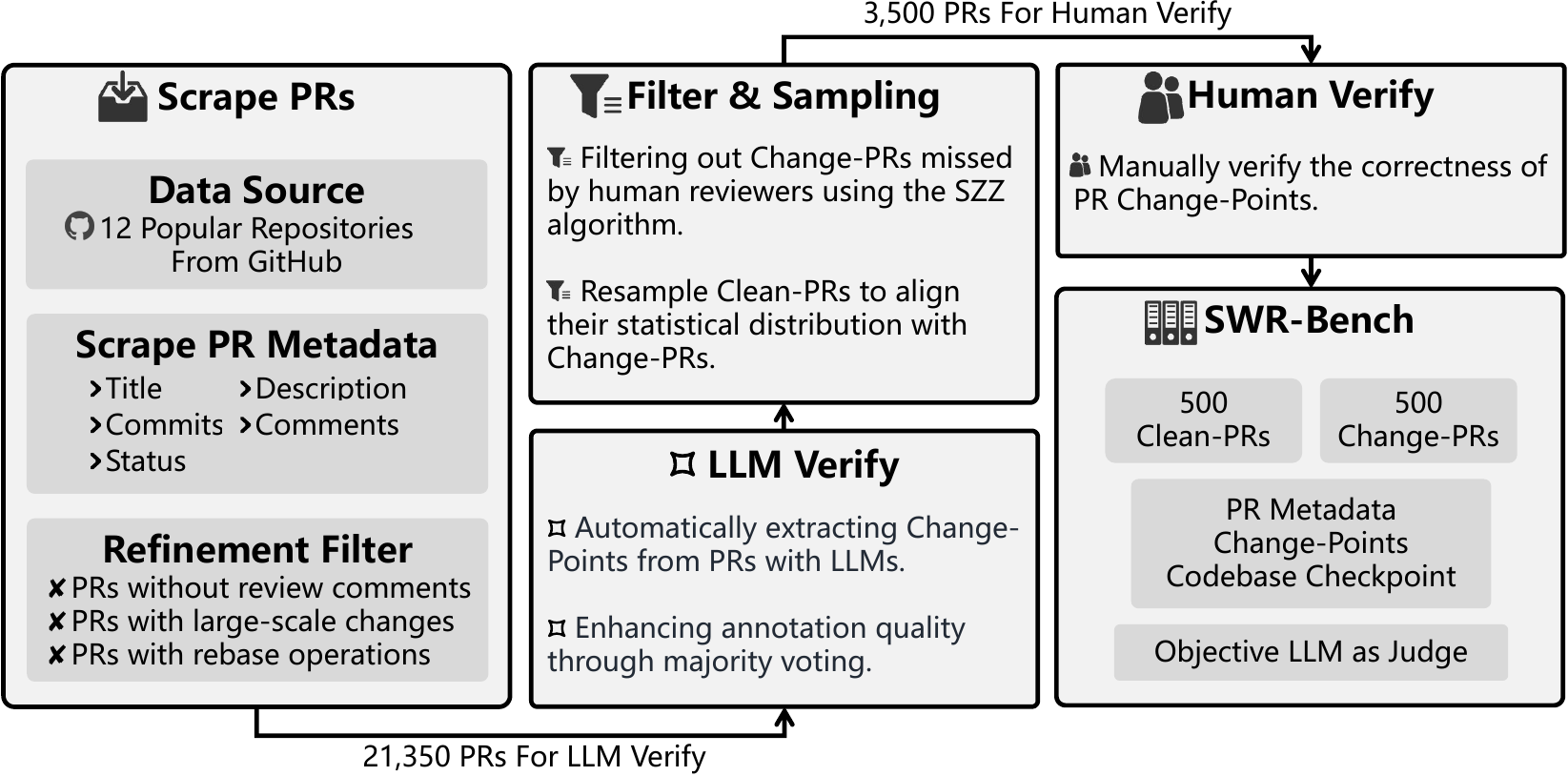} % 假设图片文件名是 swrbench.pdf
\caption{The \swrbench{} construction pipeline.}
\Description{}
\label{fig:swrbench-process}
\end{figure}

The overall construction workflow of \swrbench{} is depicted in Figure~\ref{fig:swrbench-process} and detailed below.

\subsubsection*{\textbf{Step 1: Source Data Collection and Initial Filtering}}
The foundation of \swrbench{} lies in real-world software development practices. To ensure the benchmark's realism and quality, we adopted the 12 open-source Python projects from the well-established SWE-Bench~\cite{SWE-Bench}, as they represent the most popular packages on PyPI by download count. This selection guarantees their code is of high quality and their development practices are highly representative of real-world scenarios. We utilized the GitHub API to crawl all historical pull requests (PRs) from these projects, collecting comprehensive metadata for each. This included titles, descriptions, commit histories, code diffs, review comments, discussion threads, and final PR statuses (e.g., merged, closed).

From this raw collection, we performed initial filtering to refine the candidate pool. We excluded PRs without any review comments, as they lack the necessary interaction for our analysis. We also removed PRs with large-scale changes (e.g., exceeding 10 commits in one PR) or those with rebase operations in their commit history. This is because extensive changes make it challenging for the LLM in the subsequent verification stage to correctly identify and label \cas{}, and rebase operations complicate the construction of reproducible environments. After this initial filtering, 21,350 PRs remained for the next phase.

\subsubsection*{\textbf{Step 2: LLM-based \cas{} Verification and Classification}}
In this step, we employed a LLM to identify and extract ``\cas{}'' from the filtered PRs. 
Specifically, we provided the LLM with the complete, chronologically ordered review timeline for each PR, which included its title, description, all developer-reviewer discussions, and all commit messages. A crucial instruction was for the LLM to identify each \ca{} (corresponding to the 11 types in Table~\ref{tab:change_types_condensed}) and link it to the specific commit that introduced and fixed the issue, with the full prompt is detailed in~\cite{swrbench} due to space constraints. To ensure high-quality annotations, we employed Gemini-2.5-Pro~\cite{Gemini-2.5-Pro}, a state-of-the-art (SOTA) LLM. Furthermore, we implemented a majority voting technique by making three independent requests to the LLM for each PR. A PR was considered for further processing only if all three requests yielded consistent \cas{} extraction results. 
\note{BQ1, BQ2: }\add{This requirement acts as a quality filter, excluding PRs where the LLM produced inconsistent results across runs.} 
Lastly, those consistent PRs were then classified into two categories: \changeprs{}'', which contain at least one identified \ca{}, and \cleanprs{}'', which have no identified \ca{}.

\subsubsection*{\textbf{Step 3: Quality Enhancement through Filtering and Sampling}}
To further enhance dataset quality, we performed additional filtering and sampling. First, we applied the SZZ algorithm~\cite{SZZ} to identify issues that might have been missed by human reviewers but were fixed in later commits. PRs containing such missed changes were removed. This step ensures that our \cleanprs{} are genuinely free of known, non-trivial defects. Consequently, any comment an ACR tool generates for a \cleanpr{} is, by definition, considered a false positive.
This strict protocol enables a robust measurement of the false positive rate, a critical metric for assessing a tool's precision and its ability to avoid distracting developers with irrelevant suggestions.
\note{CQ3: }\add{We acknowledge that SZZ has known limitations in precision and recall~\cite{SZZ}, as it may miss a small number of latent bugs or produce false links. We therefore employed it solely as a supplementary heuristic to improve dataset completeness, not as a sole ground-truth source, and all results were subject to subsequent manual verification (Step 4). Moreover, since all tools are evaluated on the same SZZ-augmented dataset, this still provides a fair basis for comparative analysis.}

Next, to prevent models from relying on superficial heuristics, we addressed the statistical differences between \changeprs{} and \cleanprs{}. As highlighted in prior work~\cite{deepjit}, models can exploit simple metrics (e.g., lines of code changed, number of commits) to distinguish between PRs needing changes and those that do not, without truly understanding the code. To mitigate this and create a more robust evaluation, we performed stratified sampling~\cite{StratifiedSampling} on the \cleanprs{} to align their distribution of key statistical properties with that of the \changeprs{}. While this intentionally deviates from the natural data distribution, it creates a more challenging and meaningful benchmark that forces models to engage with code logic.

\subsubsection*{\textbf{Step 4: Manual Verification and Refinement}}
After the preceding steps, approximately 3,500 PRs remained. Given the significant effort required for manual annotation, we randomly sampled 1,000 \changeprs{} and 1,000 \cleanprs{} for rigorous manual verification. 
\note{AQ1, BQ2: }\add{This process was conducted by five experienced graduate students in computer science, each holding at least a Master's degree and possessing over two years of software development experience. Specifically, we assigned 800 PRs to each annotator, organized such that each PR was independently annotated by exactly two annotators. The primary goal of this manual verification was to check the correctness of each LLM-identified \ca{} label, ensuring it conformed to our definition and could be accurately classified into one of the 11 types in Table~\ref{tab:change_types_condensed}. Additionally, annotators filtered out \changeprs{} containing only ``trivial \cas{}'', which we define as changes related solely to formatting or documentation updates (types E.1 and E.2 in Table~\ref{tab:change_types_condensed}). This was a deliberate choice to create a more challenging and realistic benchmark, as a model excelling on \swrbench{} by being forced to address substantive issues is more likely to be effective in real-world scenarios.}

\add{To ensure annotation reliability, we calculated Cohen's Kappa~\cite{CohensKappa} for each pair of independent reviews, yielding a score of 66.08. While this indicates ``substantial agreement,'' it also reflects the inherent difficulty and ambiguity of identifying specific \cas{} from PR discussions. To mitigate the impact of this ambiguity on dataset quality, all disagreements were resolved through discussion between the two assigned annotators to ensure the correctness of the final labels. This rigorous verification process was crucial for ensuring the high quality and reliability of \swrbench{}.}

\subsubsection*{\textbf{Dataset Statistics}}

The final \swrbench{} dataset consists of 500 \changeprs{} and 500 \cleanprs{}, randomly selected from manually verified and corrected PRs. This balanced composition is vital for robustly evaluating ACR tools, particularly their false positive rate, a common critique of existing automated review systems~\cite{gptreview,BitsAI-CR}.

Each \swrbench{} instance offers a comprehensive dataset for the evaluation of ACR tools, capturing the state of a PR right before its first manual review. This dataset is composed of three key components: (1) the metadata data of the PR (title, description, commits made before the first manual review), (2) ground-truth \ca{} (only for those made prior to the first manual review) information for \changeprs{}, detailing their type, description, and associated commit SHA, and (3) codebase checkpoint information, allowing for the reconstruction of the full PR codebase at the relevant commit. This latter feature is particularly important for agent-based ACR tools that require an interactive environment. 

\begin{figure}[htbp]
    \centering
    % 左侧放置表格 (给表格分配 56% 的宽度，因为表格列数较多)
    \begin{minipage}[c]{0.48\linewidth}
        \centering
        \captionof{table}{Overall statistics of the \swrbench{}. ``Avg.'' denotes average values of all instances.}
        \label{tab:swrbench_stats}
        \resizebox{\linewidth}{!}{%
        \begin{tabular}{|c|c|c|c|c|c|c|}
        \hline
        \textbf{PR Type} & \textbf{Count} & \thead{\textbf{Avg.} \\ \textbf{Commits}} & \thead{\textbf{Avg.} \\ \textbf{Files}} & \thead{\textbf{Avg.} \\ \textbf{Lines +}} & \thead{\textbf{Avg.} \\ \textbf{Lines -}} & \thead{\textbf{Avg. Change} \\ \textbf{Points}} \\ \hline\hline
        \textbf{Change-PR} & 500   & 4.05         & 6.29       & 123.64           & 60.69              & 1.90               \\ 
        \textbf{Clean-PR}  & 500   & 3.25         & 6.85       & 112.60           & 67.22              & 0.00               \\ 
        \textbf{All}       & 1000  & 3.65         & 6.57       & 118.12           & 63.96              & 0.95               \\ \hline
        \end{tabular}%
        }
    \end{minipage}\hfill
    % 右侧放置图片 (给饼图分配 40% 的宽度，中间留 4% 的弹性空白)
    \begin{minipage}[c]{0.50\linewidth}
        \centering
        \includegraphics[width=\linewidth]{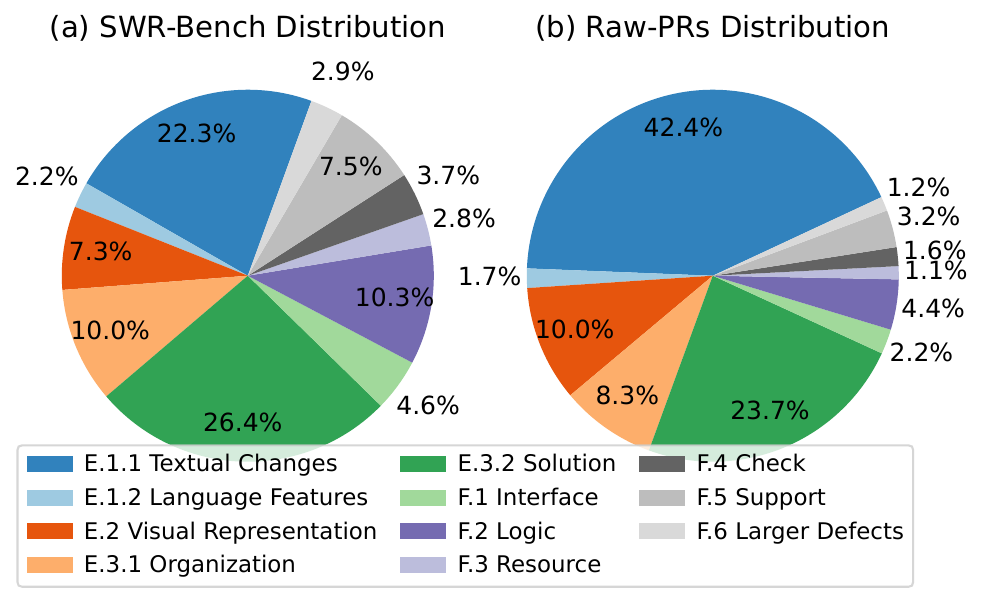}
        \captionof{figure}{Change type distribution of \swrbench{}.}
        \Description{}
        \label{fig:swrbench_distribution}
    \end{minipage}
\end{figure}

Table~\ref{tab:swrbench_stats} presents the overall dataset statistics. We find that, on average, \changeprs{} have similar numbers of modified files and lines of code compared to \cleanprs{}. These similarities underscore the effectiveness of our resampling strategy (Step 3) in ensuring that \cleanprs{} are not trivially distinguishable from \changeprs{} based on such coarse-grained metrics. 

\note{AQ4: }\add{Furthermore, we analyzed the \ca{} type distribution (Figure~\ref{fig:swrbench_distribution}). In Raw-PRs (before manual verification and filtering), functional changes (e.g., logic errors, performance issues) constituted less than 15\% of all identified changes, with a predominance of more evolutionary changes like documentation or style adjustments. After our manual verification process, which explicitly filtered out PRs with only trivial \cas{}, the proportion of functional changes in \swrbench{} increased to 31.8\%, while the proportion of simple textual changes was intentionally reduced. Additionally, earlier quality filtering steps (e.g., majority voting) also contribute to distribution shifts from real-world data. We argue that such deviations are acceptable, as our goal is to provide a fair and consistent evaluation scenario for ACR tools. A method that performs well on this intentionally difficult distribution, where simple cases are underrepresented, is highly likely to generalize well to practical scenarios.}

% \begin{table}[htbp]
% \centering
% \caption{Overall statistics of the \swrbench{}. ``Avg.'' denotes average values of all instances.}
% \label{tab:swrbench_stats}
% \begin{adjustbox}{width=0.5\linewidth}
% \begin{tabular}{|c|c|c|c|c|c|c|}
% \hline
% \textbf{PR Type} & \textbf{Count} & \thead{\textbf{Avg.} \\ \textbf{Commits}} & \thead{\textbf{Avg.} \\ \textbf{Files}} & \thead{\textbf{Avg.} \\ \textbf{Lines +}} & \thead{\textbf{Avg.} \\ \textbf{Lines -}} & \thead{\textbf{Avg. Change} \\ \textbf{Points}} \\ % MODIFIED

% \hline\hline

% \textbf{Change-PR} & 500   & 4.05         & 6.29       & 123.64           & 60.69              & 1.90               \\ % MODIFIED
% \textbf{Clean-PR}  & 500   & 3.25         & 6.85       & 112.60           & 67.22              & 0.00               \\ % MODIFIED
% % \midrule
% \textbf{All}       & 1000  & 3.65         & 6.57       & 118.12           & 63.96              & 0.95               \\ % MODIFIED
% \hline
% \end{tabular}
% \end{adjustbox}
% \end{table}

% %pie_chart_1
% \begin{figure}[htbp] 
% \centering
% \includegraphics[width=0.5\linewidth]{figures/change_type_pie_charts.pdf}
% \caption{Change type distribution of \swrbench{}.}
% \Description{}
% \label{fig:swrbench_distribution}
% \end{figure}

\subsection{Evaluation Methodology}
\label{subsec:evaluation_methodology}

Evaluating the output of ACR tools presents unique challenges. The report generated by a model is typically unstructured natural language text, containing various comments, or identified issues. Directly comparing this with a structured list of ground truth \cas{} is difficult.
As discussed (Section~\ref{sec:background}), traditional text similarity metrics fail to capture semantic accuracy and practical value in code review comments~\cite{metric_weakness,metric_weakness_2}. Consequently, we turn to solutions utilizing large language models as evaluation aids (LLM-as-Judge)~\cite{mtbench,pandalm}. However, we also recognize that traditional LLM-as-Judge methods often rely on the LLM's subjective judgment (e.g., scoring comment quality), which can introduce bias and inconsistency~\cite{LLM-as-judge,LLM-as-judge_2,LLM-as-judge_3,LLM-as-judge_4}. To mitigate this, we designed an objective LLM evaluation framework that leverages ground-truth \cas{} referencing. Instead of subjective scoring or ranking, our LLM performs a fact-based matching task: determining if issues in the model's report correspond to predefined ground truth \cas{}.

\begin{figure}[htbp]
\centering
% You will need to create this figure and name it, for example, 'evaluation_pipeline.pdf'
\includegraphics[width=0.5\linewidth]{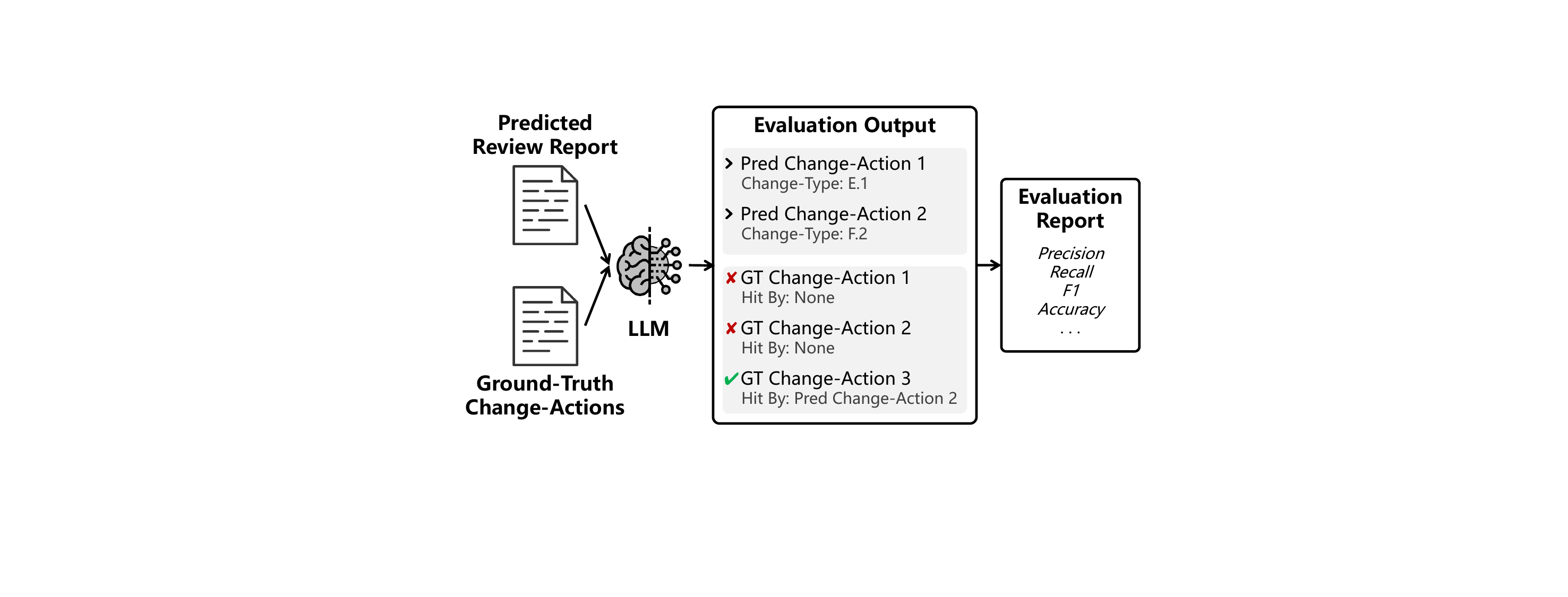} % Replace with your actual figure path
\caption{The objective LLM-based evaluation pipeline for predicted review report.}
\Description{}
\label{fig:evaluation_pipeline}
\end{figure}

Our evaluation process, depicted in Figure~\ref{fig:evaluation_pipeline}, unfolds as follows:

The inputs to the evaluation pipeline are the code review report generated by the ACR tool and the corresponding ground truth \cas{} from our \swrbench{}.
First, an evaluation LLM parses the ACR tool's unstructured report to extract distinct predicted \cas{} \note{MV2, AQ6: }\add{(a single report generated by the ACT tool contains multiple \cas{})}. For each predicted \ca{}, the LLM is also prompted to determine its \textit{change-type} (from 11 types in Table~\ref{tab:change_types_condensed}). These latter two are subjective LLM assessments but provide useful metadata about the review report.
Next, the core objective evaluation involves the LLM performing a matching task. It compares each extracted predicted \ca{} against the PR's set of ground truth \cas{}. The LLM determines, for each ground truth \ca{}, whether it has been successfully ``hit'' (i.e., semantically identified) by one or more predicted \cas{}.

Based on this explicit matching information, we can define \textit{TP}, \textit{FP}, and \textit{FN}. These definitions then enable the straightforward calculation of standard performance metrics like $Precision=\frac{TP}{TP+FP}$, $Recall=\frac{TP}{TP+FN}$, and $F1=2\times\frac{Precision\times Recall}{Precision+Recall}$:
\begin{itemize}[leftmargin=*, label=\textbullet, nosep, topsep=0pt, partopsep=0pt]
\item \textbf{\textit{True Positives (TP)}:} Ground-truth \cas{} successfully hit by at least one predicted \ca{}.
\item \textbf{\textit{False Positives (FP)}:} The number of predicted \cas{} that do not hit any ground truth \ca{}.
\item \textbf{\textit{False Negatives (FN)}:} The number of ground truth \cas{} not hit by any predicted \cas{}.
\end{itemize}

Furthermore, as discussed in Section~\ref{subsec:benchmark_construction}, functional changes constitute a critical, but smaller portion of the \cas{} in \swrbench{}. To provide a more nuanced understanding of an ACR tool's ability to detect these functional issues, we report performance metrics ($Precision$, $Recall$, and $F1$) specifically for functional changes in addition to the overall metrics. This specialized calculation considers only predicted \cas{} that the evaluation LLM has typed as functional change and ground-truth \cas{} with functional type.

% In summary, our key evaluation metrics include overall \textbf{$Precision$}, \textbf{$Recall$}, and \textbf{$F1$} for all \cas{}, and the specific \textbf{$Precision$}, \textbf{$Recall$}, and \textbf{$F1$} for functional \cas{}. We also report descriptive statistics such as the average number of predicted \cas{} per PR for the predicted report. 
This objective, ground-truth-referenced LLM evaluation approach allows for a more accurate and in-depth assessment of ACR tools, overcoming traditional method limitations. The reliability of this evaluation approach is further validated through manual verification in later experiments.

\section{Evaluation and Study}
\label{sec:evaluation_and_study}
% In this section, we first establish the effectiveness of our proposed evaluation methodology. Subsequently, we conduct extensive experiments on \swrbench{} to demonstrate its contributions and utility in assessing automated code review tools.

\subsection{Research Questions}
Our study is guided by the following research questions:
\label{subsec:research_questions}
\begin{itemize}[leftmargin=*, nosep]
    
    \item \textbf{\textit{RQ1: How reliable is the proposed evaluation methodology?}}
    % In this RQ, we validate the reliability of our proposed evaluation methodology by verifying the consistency between its results and human evaluation results.
    \item \textbf{\textit{RQ2: How do mainstream automated code review tools and LLMs perform on \swrbench{}?}}
    % In this RQ, we conduct a comprehensive performance analysis of existing mainstream ACR tools and SOTA LLMs on the \swrbench{} benchmark.
    \item \textbf{\textit{RQ3: How can the performance of automated code review tools be improved?}}
    % In this RQ, we investigate feasible strategies to enhance the code review performance of LLM-based ACR tools.
\end{itemize}

\subsection{Subjects of Study}
\label{subsec:subjects_of_study}
\subsubsection{Large Language Models (LLMs)}
\label{ssubsec:llms}
To comprehensively evaluate current LLMs on ACR tasks, we selected a diverse range of models, including closed-source LLMs like the GPT series (o3, o4-mini, GPT-4o, GPT-5), Gemini series (Gemini-2.5-Pro, Gemini-2.5-Flash), and Claude series (Claude-3.7-Sonnet, Claude-4-Sonnet, Claude-4-Opus). We also include open-source LLMs such as DeepSeek series (DeepSeek-R1, DeepSeek-V3) and Qwen2.5 series (Qwen2.5-Chat-7B/14B/32B, Qwen2.5-R1-7B/14B/32B).
% This diverse set allows assessment across the current proprietary and public technology landscape. Importantly, by including base and reasoning-enhanced versions for some models (e.g., Qwen2.5-Chat and Qwen2.5-R1), we can preliminarily investigate if reasoning enhancements directly benefit complex code review tasks.

\subsubsection{Automated Code Review (ACR) Baselines}
For a comprehensive evaluation, we benchmark against representative ACR tools and methodologies:
\begin{itemize}[leftmargin=*, nosep]
    \item \textbf{LLM-Review} (Prompting Baseline): Our straightforward baseline that directly feeds PR metadata and diffs to an LLM via a simple prompt, measuring vanilla review capabilities.
    \item \textbf{SWR-Agent} (Agentic Baseline): Inspired by SWE-Agent~\cite{SWE-agent}, we built this baseline to adapt agentic workflows for code review, allowing the LLM to use tools, explore the codebase, and gather context.
    \item \textbf{CR-Agent}~\cite{CodeAgent}: A multi-agent tool using two interacting agents (focusing on formatting and functional defects, respectively) that discuss and debate to produce a final report.
    \item \textbf{Hybrid-Review}~\cite{Hybrid-Code-Review}: Extends the simple prompting baseline by additionally injecting static analysis reports as supplementary context.
    \item \textbf{PR-Review}~\cite{PRReview}: While similar to LLM-Review, PR-Review uses more sophisticated prompt engineering. Specifically, it first consolidates code diffs from multiple commits to form the input. When faced with context length limitations, it prioritizes files based on importance, potentially excluding less critical files. Finally, it instructs the LLM to output the review report in a specific structured format, categorizing suggestions into areas such as test-related defects, security-related defects, and other areas for improvement.
    \item \textbf{Code-Reviewer}~\cite{codereviewer}: A traditional hunk-level method using a fine-tuned CodeT5 model~\cite{codet5}. We adapt our PRs into diff hunks for its input and concatenate its generated hunk-level comments to form a full PR review.
    \item \textbf{Llama-Reviewer}~\cite{acr_llm_4}: Fine-tunes a Llama-Base-7B model~\cite{llama1} for hunk-level review. We evaluate it using the same hunk-to-PR concatenation protocol as Code-Reviewer.
\end{itemize}
Lastly, to ensure fair comparison, all evaluated ACR tools used their official default configurations.

% \subsection{Results and Analysis}

\subsection*{\textbf{RQ1: How reliable is the proposed evaluation methodology?}}

\begin{figure*}[htbp] 
\centering
\includegraphics[width=0.80\linewidth]{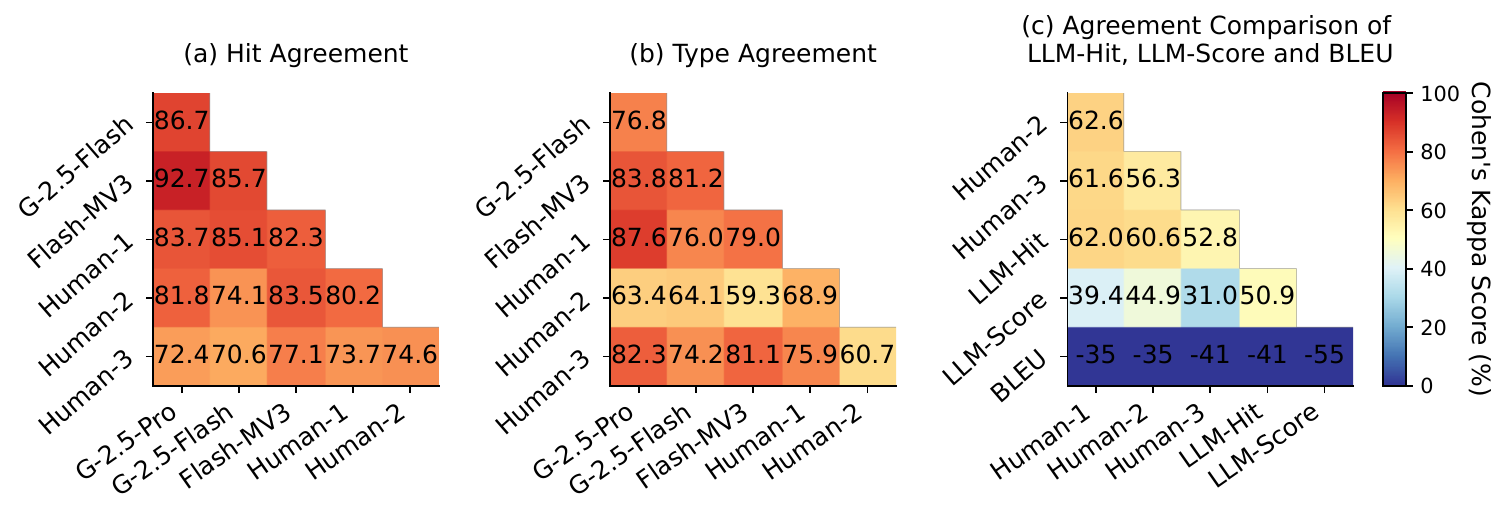}
\caption{Validation of the proposed objective evaluation methodology. \note{BQ5: }\add{(a) and (b) show the inter-rater agreement (Cohen's Kappa) for \textbf{Hit} identification and \textbf{Type} classification, respectively.} \note{MV1, AQ2: }\add{(c) compares the agreement (Cohen's Kappa) of two LLM-as-Judge methodology and BLEU judge methodology with human preferences in a pairwise comparison task.}}
\Description{}
\label{fig:heatmap_combined}
\end{figure*}

To rigorously assess the reliability of our proposed LLM-as-Judge methodology, we conducted two validation experiments.

First, we measured inter-rater and inter-model agreement for our core evaluation tasks. We randomly selected 100 code review reports from RQ2 and had them independently annotated by three human experts and three distinct LLM judges (Gemini-2.5-Pro, Gemini-2.5-Flash and Gemini-2.5-Flash with majority voting ($n$=3)). We measured agreement using Cohen's Kappa for identifying ``Hit'' of each ground-truth \cas{} and classifying their ``Type''.

\textit{Hit Agreement} (Figure~\ref{fig:heatmap_combined}a): The agreement for our primary "Hit" metric was exceptionally high, with Kappa scores ranging from 70.6 to 86.7 across all human-human, human-LLM, and LLM-LLM pairs. This indicates almost perfect agreement and confirms that identifying "Hits" is an objective task. The high consistency between Gemini-2.5-Pro and Gemini-2.5-Flash (Kappa = 86.7) demonstrates that the evaluation is stable across different models, implying that the results are not sensitive to the choice of a specific LLM judge, and therefore are unlikely to be sensitive to the randomness of multiple runs.
\note{BQ5: }\add{Furthermore, the majority voting variant achieved Kappa scores comparable to the single-call Gemini-2.5-Flash and Gemini-2.5-Pro, indicating that majority voting provides only marginal improvement for this already highly consistent task.}

\textit{Type Agreement} (Figure~\ref{fig:heatmap_combined}b): For "Type" classification, human-LLM kappa scores ranged from 63.4 to 87.6. The agreement between LLMs and humans was comparable to the agreement among humans themselves (human-human Kappa: 60.7 to 75.9). This suggests that the LLM judge's performance on Type classification is consistent with that of human experts.
\note{BQ5: }\add{Similarly, the majority voting variant of Gemini-2.5-Flash showed no significant improvement over the single-call version for Type classification, further confirming that a single LLM call is sufficient for reliable evaluation.}

Second, to demonstrate the superiority of our objective methodology over traditional approaches, we compared it against several alternative evaluation methods. We randomly selected 1,000 pairs of review reports from RQ2, where each pair containing two distinct reviews for the same PR, and asked three human experts to choose the better report in each pair. We then tasked the following automated methods (both using Gemini-2.5-Flash) to perform the same comparison:
1) \textit{LLM-Hit-Judge} (ours), our proposed method, which prefers the report that achieves more ``Hits'' on ground-truth \cas{} with fewer false positives.
2) \textit{LLM-Score-Judge}, a traditional approach where the LLM assigns a holistic quality score (1-10) to each report, with the higher-scoring report being preferred.
\note{MV1, AQ2: }\add{3) \textit{Bleu}~\cite{BLEU} and \textit{Rouge-L}~\cite{rouge}, traditional text similarity metrics that compare the generated review report against the ground-truth review discussion text, with the report achieving a higher score being preferred.}

As shown in Figure~\ref{fig:heatmap_combined}c, \textit{LLM-Hit-Judge} achieved substantial agreement with human preferences, with Cohen's Kappa scores ranging from 52.8 to 62.0. This level of agreement is on par with the inter-human agreement (Kappa range: 56.3 to 62.6). In contrast, the baseline \textit{LLM-Score-Judge} showed significantly lower agreement with humans (Kappa range: 31.0 to 44.9), highlighting the unreliability of subjective scoring. 
\note{MV1, AQ2: }\add{Most notably, the traditional text similarity metrics performed drastically worse. \textit{BLEU} achieved Kappa scores ranging from only -35 to -41 with human preferences, indicating agreement even worse than chance. This result validates that our objective, change-action-based evaluation aligns more closely with expert human judgment.}

Ultimately, Gemini-2.5-Pro, Gemini-2.5-Flash and Majority-Voting demonstrated commendable and similar agreement levels with human evaluators, particularly for the primary ``Hit'' metric. Given that Gemini-2.5-Flash demonstrated strong, reliable performance comparable to Gemini-2.5-Pro at a significantly lower cost (approx. \$1.57 for a full SWR-Bench evaluation), we selected it for all subsequent experiments.

\mybox{Conclusion 1}{Our objective, change-action-based evaluation methodology is highly reliable, stable across different LLM judges (with or without majority voting), and demonstrates superior alignment with human expert judgment compared to both traditional subjective scoring methods and text similarity metrics.}

\subsection*{\textbf{RQ2: How do mainstream automated code review tools and LLMs perform on \swrbench{}?}}

% Please add the following required packages to your document preamble:
% \usepackage{multirow}
% \usepackage{graphicx}
\begin{table*}[ht]
\centering
\caption{Evaluation of studied ACR approaches on \swrbench{}. The table shows hit-based \textit{Precision}, \textit{Recall}, and \textit{F1}. \textit{Avg. Count} indicate the average number of predicted \cas{}, while \textit{Avg. FP Count} indicate the average false positive number. \note{AQ2, BQ3: Add text similarity metrics and false positive metrics.}}
\label{tab:rq2_performance_summary}
\resizebox{1\textwidth}{!}{%
\begin{tabular}{|ll|ccccc|ccccc|cc|}
\hline
\multicolumn{1}{|l|}{\multirow{3}{*}{\textbf{ACR Tools}}} & \multicolumn{1}{l|}{\multirow{3}{*}{\textbf{LLMs}}} &  \multicolumn{5}{c|}{\textbf{Overall \CAS{}}}                                                                                                                                   & \multicolumn{5}{c|}{\textbf{Functional \CAS{}}} & \multicolumn{2}{c|}{\textbf{Text Similarity}}                                                                                                                               \\ \cline{3-14}
\multicolumn{1}{|l|}{} &  & \textbf{\textit{Precision}} & \textbf{\textit{Recall}} & \textbf{\textit{F1}} & \textbf{\begin{tabular}[c]{@{}c@{}}\textit{Avg.}\\ \textit{Count}\end{tabular}} & \textbf{\begin{tabular}[c]{@{}c@{}}\textit{Avg. FP}\\ \textit{Count}\end{tabular}} & \textbf{\textit{Precision}} & \textbf{\textit{Recall}} & \textbf{\textit{F1}} & \textbf{\begin{tabular}[c]{@{}c@{}}\textit{Avg.}\\ \textit{Count}\end{tabular}} & \textbf{\begin{tabular}[c]{@{}c@{}}\textit{Avg. FP}\\ \textit{Count}\end{tabular}} & \textbf{\textit{Bleu}} & \textbf{\textit{Rouge-L}}
\\ \hline\hline
\multicolumn{1}{|l|}{\multirow{4}{*}{\textbf{LLM-Review}}} & \textbf{Gemini-2.5-Pro} & 8.02 & 21.29 & 11.65 & 2.47 & 2.28 & 22.60 & 26.86 & 24.54 & 0.35 &0.27& 0.58 &	9.64 \\
\multicolumn{1}{|l|}{} & \textbf{Claude-3.7-Sonnet} & 9.85 & 14.31 & 11.67 & 1.35 &1.22& \textbf{24.37} & 16.02 & 19.33 & 0.20 &0.15& 0.47	& 10.53\\
\multicolumn{1}{|l|}{} & \textbf{DeepSeek-R1} & 9.79 & 25.58 & 14.16 & 2.44 &2.20& 15.53 & 32.02 & 20.92 & 0.61 &0.52& 0.50 & 10.19\\
\multicolumn{1}{|l|}{} & \textbf{Mean} & 9.22 & 20.39 & 12.49 & 2.09 &1.90& 20.83 & 24.97 & 21.60 & 0.39 &0.31& 0.52 &	10.12\\ \hline\hline
\multicolumn{1}{|l|}{\multirow{3}{*}{\textbf{SWR-Agent}}} & \textbf{Gemini-2.5-Pro} & 9.93 & 19.14 & 13.07 & 1.80 &1.62& 18.11 & 25.84 & 21.30 & 0.42 &0.35& 1.95 & 11.45\\
\multicolumn{1}{|l|}{} & \textbf{Claude-3.7-Sonnet} & 8.29 & 22.72 & 12.15 & 2.55 &2.34& 17.92 & 30.90 & 22.68 & 0.51 &0.42& 1.46	& 11.77\\
\multicolumn{1}{|l|}{} & \textbf{Mean} & 9.11 & 20.93 & 12.61 & 2.18 &1.98& 18.02 & 28.37 & 21.99 & 0.47 &0.38& 1.31 &	11.11\\ \hline\hline
\multicolumn{1}{|l|}{\multirow{4}{*}{\textbf{CR-Agent}}} & \textbf{Gemini-2.5-Pro} & 6.97 & 17.63 & 9.98 & 2.35 &2.18& 16.26 & 26.55 & 20.17 & 0.48 &0.40& 0.74 &	7.43\\
\multicolumn{1}{|l|}{} & \textbf{Claude-3.7-Sonnet} & 6.30 & 17.17 & 9.21 & 2.54 &2.38& 18.06 & 22.78 & 20.15 & 0.38 &0.31& 1.00	& 10.22\\
\multicolumn{1}{|l|}{} & \textbf{DeepSeek-R1} & 5.42 & 19.50 & 8.48 & 3.36 &3.17& 11.45 & 29.44 & 16.49 & 0.77 &0.68& 1.00 & 9.87\\
\multicolumn{1}{|l|}{} & \textbf{Mean} & 6.23 & 18.10 & 9.22 & 2.75 &2.58& 15.26 & 26.26 & 18.94 & 0.54 &0.47& 0.91 & 9.17\\ \hline\hline
\multicolumn{1}{|l|}{\multirow{4}{*}{\textbf{Hybrid-Review}}} & \textbf{Gemini-2.5-Pro} & 2.83 & 20.44 & 4.97 & 6.64 &6.42& 11.09 & 29.28 & 16.08 & 0.80 &0.71& 0.77 & 7.28\\
\multicolumn{1}{|l|}{} & \textbf{Claude-3.7-Sonnet} & 2.20 & 11.23 & 3.68 & 4.63 &4.52& 7.95 & 15.64 & 10.55 & 0.59 &0.54& 0.39	& 5.30\\
\multicolumn{1}{|l|}{} & \textbf{DeepSeek-R1} & 3.33 & \textbf{28.44} & 5.96 & \textbf{7.95} &\textbf{7.60}& 10.10 & 47.31 & 16.65 & \textbf{1.47} &\textbf{1.31}& 1.02 & 8.41\\
\multicolumn{1}{|l|}{} & \textbf{Mean} & 2.79 & 20.04 & 4.87 & 6.41 &6.18& 9.71 & 30.74 & 14.43 & 0.95 &0.85& 0.73 &	7.00\\ \hline\hline
\multicolumn{1}{|l|}{\multirow{4}{*}{\textbf{PR-Review}}} & \textbf{Gemini-2.5-Pro} & \textbf{16.65} & 23.18 & \textbf{19.38} & 1.32 &1.10& 19.38 & 40.72 & \textbf{26.26} & 0.65 &0.52& 0.40 & 8.92\\
\multicolumn{1}{|l|}{} & \textbf{Claude-3.7-Sonnet} & 14.90 & 23.50 & 18.23 & 1.50 &1.27& 14.72 & 40.32 & 21.56 & 0.86 &0.74& 0.36	& 8.38\\
\multicolumn{1}{|l|}{} & \textbf{DeepSeek-R1} & 14.61 & 25.50 & 18.58 & 1.66 &1.41& 15.55 & \textbf{50.62} & 23.79 & 1.06 &0.89& 0.31	& 7.70\\
\multicolumn{1}{|l|}{} & \textbf{Mean} & 15.39 & 24.06 & 18.73 & 1.49 &1.26& 16.55 & 43.89 & 23.87 & 0.86 &0.72& 0.36 & 8.34\\ \hline\hline
\multicolumn{2}{|l|}{\textbf{Code-Reviewer}} & 4.19 & 11.35 & 6.13 & 2.55 &2.44& 9.46 & 14.67 & 11.50 & 0.47 &0.42& \textbf{10.25} & 22.14\\ \hline\hline
\multicolumn{2}{|l|}{\textbf{Llama-Reviewer}} & 4.28 & 23.08 & 7.22 & 4.91 &4.70& 8.62 & 37.02 & 13.98 & 1.32 &1.20& 7.38 & \textbf{22.89}\\ \hline
\end{tabular}%
}
\end{table*}

In this RQ, we first investigate the performance of studied ACR approaches on \swrbench{}, and each approaches was evaluated using three powerful LLMs, with the exception of SWR-Agent, which was not run with DeepSeek-R1 due to the latter's lack of function call capabilities. For each ACR technique, we also report the mean performance metrics across the LLMs it was paired with. The detailed results are summarized in Table~\ref{tab:rq2_performance_summary}. % <--- ASSUMING A TABLE WILL SUMMARIZE THESE NUMBERS

The results in Table~\ref{tab:rq2_performance_summary} show that SOTA ACR techniques, when paired with SOTA LLMs, are not yet ready for real-world code review deployment based on their performance on \swrbench{}. For instance, the top-performing combination, PR-Review leveraged with Gemini-2.5-Pro, achieved an $F1$ score of only 19.38\%. A primary factor limiting higher $F1$ scores for all techniques is their low precision, indicative of a high false positive rate. \add{Specifically, the \textit{Avg. FP Count} metric reveals that these approaches frequently generate multiple invalid suggestions per pull request, with some combinations (e.g., Hybrid-Review with DeepSeek-R1) producing over 7 false positives on average.} This issue is more severe for other four ACR techniques, all of which exhibited precision scores below 10\%. This implies that while these ACR techniques can identify some valuable issues (or ``\cas{}''), their overall effectiveness is severely undermined by an excessive number of false positives. Consequently, developers would need to invest considerable additional effort in verifying the validity of the generated reports, hindering practical adoption. This observation aligns with findings from previous empirical studies~\cite{gptreview} and further validates the efficacy of \swrbench{} as a benchmark that closely mirrors realistic code review scenarios.

\mybox{Conclusion 2}{Current ACR approaches, even when augmented with advanced LLMs, demonstrate limited performance on the \swrbench{}, primarily constrained by high false positive rates. This significantly hinders their immediate applicability in practical code review workflows.}

\paragraph{Performance Analysis by ACR Tools}
A granular analysis of $F1$ scores reveals distinct performance tiers among the ACR tools (Table~\ref{tab:rq2_performance_summary}). PR-Review, leveraging meticulous prompt engineering, achieved the highest overall performance ($Overall\text{-}F1$: 18.73\%). This significantly surpassed both SWR-Agent and LLM-Review (approx. 12\%). 

Meanwhile, other approaches showed significant limitations. The multi-agent CR-Agent performed poorly ($Overall\text{-}F1$: 9.22\%), likely due to interaction overhead and error propagation—known challenges in multi-agent systems~\cite{multiagentfail}. The worst performer, Hybrid-Review ($Overall\text{-}F1$: 4.87\%), suffered from extremely low precision (with the highest mean \textit{Avg. FP Count} of 6.18), indicating that simply integrating raw static analysis outputs is ineffective without careful processing.

Furthermore, the poor performance of traditional fine-tuned models (Code-Reviewer: 6.13\% $F1$, Llama-Reviewer: 7.22\% $F1$) is also notable. It highlights a critical flaw in prior approaches: models optimized for isolated code hunks fail at the holistic, contextual task of reviewing a full pull request. This underscores the value of \swrbench{} in providing a more realistic evaluation and guiding research toward end-to-end solutions.

\note{MV1, AQ2, BQ3: }\add{Table~\ref{tab:rq2_performance_summary} also reports \textit{Bleu}~\cite{BLEU} and \textit{Rouge-L}~\cite{rouge} scores. Interestingly, text similarity inversely correlates with $F1$: hunk-level models achieve high \textit{Bleu} but low $F1$, whereas PR-level tools (e.g., PR-Review) show the opposite. Our manual analysis reveals this stems from formatting differences rather than review quality. PR-Review generates structured, detailed reports that differ from brief, colloquial human ground truths, yielding low \textit{Bleu} despite capturing actual issues (high $F1$). Conversely, hunk-level models mimic human chat, often generating short, ambiguous comments that inflate \textit{Bleu} but fail to identify actionable defects (low $F1$). Consistent with RQ1 (Figure~\ref{fig:heatmap_combined}c), this format bias confirms that \textit{Bleu} and \textit{Rouge-L} are unreliable proxies for review quality, validating our change-action hit-based evaluation.}

Lastly, comparing PR-Review and LLM-Review, both of which operate via a single-turn interaction with the LLM, PR-Review's superior performance can be attributed to its more sophisticated prompt engineering, these refinements appear to significantly reduce false positives, thereby enhancing precision and $F1$ score. 

Collectively, these results underscore that refined prompt engineering, as demonstrated by PR-Review, is currently the most effective strategy for improving ACR capabilities. Agent-based approaches require further research to overcome their architectural challenges and unlock their potential. The failure of models trained on isolated code hunks validates the necessity of holistic, end-to-end benchmarks like \swrbench{} to guide the field towards more practical solutions.

\mybox{Conclusion 3}
{PR-Review, leveraging meticulous prompt engineering, achieves the best performance on \swrbench{}, underscoring the promise of prompt engineering for improving efficacy.}

\begin{table*}[htbp] % 双栏跨栏显示。如果是单栏请改成 \begin{table}[htbp]
    \centering
    
    % 左侧表格：Multi-Change 分析
    \begin{minipage}[t]{0.40\linewidth}
        \centering
        \caption{\note{MV2, AQ6: }\add{PR-Review (Gemini-2.5-Pro) performance vs. number of ground-truth changes ($N$).}}
        \label{tab:multi_change_analysis}
        % 增加一点垂直间距让两边表格对齐更美观
        \vspace{0.15cm} 
        \resizebox{\linewidth}{!}{%
        \begin{tabular}{|c|rccc|}
        \hline
        \textbf{$N$} & \textbf{\textit{Count}} & \textbf{\textit{Precision}} & \textbf{\textit{Recall}} & \textbf{\textit{F1}} \\
        \hline\hline
        1  & 266 & 30.18 & 38.35 & 33.77 \\
        2  & 139 & 35.05 & 24.46 & 28.81 \\
        3  & 56  & 34.18 & 16.07 & 21.86 \\
        4  & 17  & 29.63 & 11.76 & 16.84 \\
        5+ & 22  & 44.12 &  8.88 & 14.78 \\
        \hline
        \end{tabular}%
        }
    \end{minipage}\hfill % \hfill 用于在两个 minipage 之间自动填充空白区
    % 右侧表格：Change Type 分析
    \begin{minipage}[t]{0.50\linewidth}
        \centering
        \caption{Performance across different change types on \swrbench{}.}
        \label{tab:change_type_performance}
        \resizebox{\linewidth}{!}{%
        \begin{tabular}{|l|cccc|}
        \hline
        \textbf{Change Type}               & \textbf{\textit{Precision}} & \textbf{\textit{Recall}} & \textbf{\textit{F1}} & \thead{\textit{\textbf{Avg.}}\\\textit{\textbf{Count}}} \\ \hline\hline
        \textbf{E.1.1 Textual Changes}     & 15.85              & 13.03           & 14.30       & 0.17                                  \\
        \textbf{E.1.2 Language Features}   & 8.28               & 7.59            & 7.85        & 0.02                                 \\
        \textbf{E.2 Visual Representation} & 12.21              & 4.39            & 6.05        & 0.02                                 \\
        \textbf{E.3.1 Organization}        & 24.94              & 12.56           & 16.45       & 0.05                                \\
        \textbf{E.3.2 Solution Approach}   & 12.66              & 19.15           & 15.21       & \textbf{0.37}                                   \\ \hline\hline
        \textbf{F.1 Interface}             & 16.97              & 40.83           & 23.55       & 0.12                                  \\
        \textbf{F.2 Logic}                 & 17.28              & \textbf{54.60}           & 26.20       & 0.35                                   \\
        \textbf{F.3 Resource}              & 15.89              & 53.45           & 24.26       & 0.10                                   \\
        \textbf{F.4 Check}                 & 13.36              & 39.70           & 19.60       & 0.12                                   \\
        \textbf{F.5 Support}               & 15.95              & 35.80           & 21.74       & 0.16                                   \\
        \textbf{F.6 Larger Defects}        & \textbf{53.31}              & 18.88           & \textbf{27.65}       & 0.01                                   \\ \hline
        \end{tabular}%
        }
    \end{minipage}
\end{table*}

\add{\paragraph{\note{MV2, AQ6: }Performance in Multi-Change PR Scenarios}
A distinctive feature of \swrbench{} is that each Change-PR contains multiple ground-truth change-actions (on average 1.90 per PR, as shown in Table~\ref{tab:swrbench_stats}). To understand how baselines perform in such scenarios, we analyzed the relationship between the number of ground-truth change-actions ($N$) in a PR and the ACR tool's performance. As detailed in Table~\ref{tab:multi_change_analysis} (using PR-Review with Gemini-2.5-Pro as a representative example), we observed that as the number of change-actions per PR increases, the Recall tends to decrease sharply (from 38.35\% for $N=1$ to 8.88\% for $N \ge 5$), indicating that tools struggle to identify all issues in more complex PRs. Conversely, Precision remains relatively stable (fluctuating roughly between 29.63\% and 44.12\%), suggesting that the quality of individual predictions does not degrade significantly with PR complexity. This analysis highlights that multi-change PRs remain a significant challenge.} % while tools can identify some issues in complex PRs, achieving comprehensive coverage of all change-actions requires further research into methods that can reason about multiple, potentially interrelated, issues within a single review pass. Our evaluation protocol directly captures this challenge by calculating per-PR Precision, Recall, and F1 based on the hit-matching between the complete set of ground-truth change-actions and the full set of predicted change-actions in each review report.}

\mybox{Conclusion 4}
{\add{ACR tools struggle to comprehensively review complex PRs. As the number of issues in a PR increases, ACR tools experience a sharp drop in Recall while Precision remains stable, highlighting their inability to comprehensively review multi-change PRs.}}

\paragraph{Performance Analysis by Change Type}

To gain a deeper understanding of how ACR tools handle different types of \cas{}, we further analyzed the average performance of PR-Review across various change types defined in Table~\ref{tab:change_types_condensed}. The results, disaggregated by change type, are shown in Table~\ref{tab:change_type_performance}. % <--- ASSUMING A FIGURE OR TABLE
A key observation is that PR-Review demonstrates significantly stronger detection capabilities for functional \cas{} compared to evolutionary \cas{}. Specifically, for the F.2 Logic change, PR-Review achieved an $F1$ score of 26.20\%, with most functional change types also yielding $F1$ scores above 21\%. In stark contrast, the highest $F1$ score for an evolutionary change type, E.3.1 Organization, was merely 16.45\%.

We hypothesize that this disparity arises from the inherent nature of these change categories. Evolutionary changes often represent optional or stylistic improvements, where the criteria for what constitutes a necessary change can vary significantly between human reviewers~\cite{pr_change_taxonomy_14} and LLMs. For instance, with E.3.2 Solution Approach, which was the most prevalent evolutionary change type, one human reviewer might suggest an alternative implementation they deem superior, while another might find the current approach acceptable. Consequently, accurately detecting such optional evolutionary changes presents a considerable challenge. 

Based on these findings, we recommend that future development of ACR tools prioritize enhancing the detection capabilities for functional changes to improve the precision for functional \cas{}. Furthermore, ACR tools could consider presenting reports for functional and evolutionary changes separately. %Such an approach could significantly improve the usability and perceived value of automated code review reports by focusing developers' attention on more critical and less subjective issues.

\mybox{Conclusion 5} % Renumbered Conclusion
{ACR tools demonstrate better performance in detecting functional changes, which likely due to the subjective nature of many evolutionary changes. Accordingly, future ACR tools could prioritize robust detection of functional changes to enhance practical utility.}

% Category analysis: {'Lack of Domain Knowledge and Conventions': 13, 'Lack of Contextual Understanding': 48, 'Over-sensitivity to Refactoring and Modification': 17, 'Vague and Unactionable Feedback': 16, 'Misjudgment of "Anti-best-practices"': 3, 'Other': 3}

% \paragraph{False Positive Analysis}

% Please add the following required packages to your document preamble:
% \usepackage{multirow}
% \usepackage{graphicx}
\begin{table}[htbp]
\centering
\caption{Categories and Proportions of Common False Positives}
\label{tab:fp_categories}
\footnotesize
\resizebox{0.82\textwidth}{!}{%
\begin{tabularx}{\linewidth}{
    | >{\centering\arraybackslash}m{3.0cm} % 第1列：水平居中 & 垂直居中
     >{\raggedright\arraybackslash}X % 第2列：现在是左对齐 & 垂直居中 (因为上面的重定义)
     >{\centering\arraybackslash}m{1.4cm} | % 第3列：水平居中 & 垂直居中
}
\hline
\textbf{Categories} & \textbf{Description} & \textbf{Proportion} \\ \hline\hline
\textbf{Lack of Contextual Understanding} & The tool applies programming rules rigidly and in isolation, ignoring crucial logical consistency established by the surrounding code within the same project or commit. & 48\% \\ \hline
\textbf{Over-sensitivity to Modification} & The tool lacks an understanding of the developer's intent and tends to treat any large-scale code modification or refactoring as a potential risk. & 17\% \\ \hline
\textbf{Vague and Unactionable Feedback} & The review feedback provided by the tool is often too broad and generic, lacking specific, actionable steps. & 16\% \\ \hline
\textbf{Lack of Domain Knowledge} & The tool fails to recognize specialized coding conventions specific to a project or technical domain (i.e., domain knowledge), causing it to flag correct and idiomatic code as anomalous. & 13\% \\ \hline
\textbf{Misjudgment of "Anti-best-practices"} & The tool relies on superficial heuristics and makes incorrect judgments when developers intentionally deviate from conventional best practices to achieve higher-level goals like test effectiveness or readability. & 3\% \\ \hline
\textbf{Other} & A catch-all for other rare causes. & 3\% \\ \hline
\end{tabularx}%
}
\end{table}

\paragraph{False Positive Analysis}

While our previous analysis highlighted the practical utility of ACR tools is critically undermined by the generation of false positives, we therefore performed a qualitative analysis to diagnose the root causes of these inaccuracies. \note{AQ8: }\add{We manually inspected 100 functional false positives, randomly sampled from a total of 1,101 false positives generated by PR-Review (based on Gemini-2.5-Pro), categorizing them as summarized in Table~\ref{tab:fp_categories}.}

\note{AQ9: }\add{The taxonomy was derived through an iterative process based on Grounded Theory~\cite{multiagentfail}. Three authors independently examined and labeled each of the 100 sampled false positives, assigning initial descriptive codes to characterize the root cause. The authors then collaboratively discussed and reconciled their labels through multiple rounds of discussion, iteratively merging, splitting, and refining categories until a stable set of themes emerged. This process resulted in the five major categories and one ``Other'' category presented in Table~\ref{tab:fp_categories}.}

As the table shows, the most prevalent issue is a \textit{Lack of Contextual Understanding} (48\%). These errors typically arise from the LLM's insufficient contextual reasoning, leading to a misunderstanding of the code. For example, in \codein{astropy/astropy-1199}, the tool warned that parameter unpacking (\codein{*result}) could raise a \codein{TypeError}, but it failed to recognize that the variable had been explicitly constructed as a tuple within the same commit, thus ensuring the operation's safety.

Other significant sources of error are \textit{Over-sensitivity to Modification} (17\%) and \textit{Vague and Unactionable Feedback} (16\%). These stem from the model's cautious bias, leading to low-value suggestions like "verifying" every new logic in \codein{scikit-learn/scikit-learn-25025}, offering no specific flaw or actionable advice. Such issues could be mitigated through prompt engineering.

Additionally, a \textit{Lack of Domain Knowledge} (13\%) contributes to false positives when the LLM is unaware of project-specific conventions. In \codein{sympy/sympy-17801}, the idiomatic expression is \codein{S.true}, used for checking symbolic boolean values, was incorrectly flagged, even though it is the standard practice in the library to avoid a \codein{TypeError}. This type of error could potentially be addressed by providing additional domain knowledge in the prompt.

In summary, these findings demonstrate that false positives arise from a fundamental inability to comprehend developer intent, code context, and domain-specific knowledge. Future work must therefore enhance the reasoning abilities of LLMs and optimize prompts to generate more accurate, actionable feedback.

\mybox{Conclusion 6} % Renumbered Conclusion
{Current ACR tools struggle to grasp three critical elements: developer intent, code context, and domain knowledge. Addressing this requires advancing LLMs' core reasoning capabilities and employing better prompting strategies to improve the quality of review reports.}

% Please add the following required packages to your document preamble:
% \usepackage{multirow}
% \usepackage{graphicx}
\begin{table*}[htbp]
\centering
\caption{LLMs performance evaluation for code review with PR-Review on \swrbench{}.}
\label{tab:llm_performance_pr_agent}
\resizebox{0.80\textwidth}{!}{%
\begin{tabular}{|ll|cccc|cccc|}
\hline
\multicolumn{2}{|c|}{\multirow{3}{*}{\textbf{LLMs}}} & \multicolumn{4}{c|}{\textbf{Overall \CAS{}}} & \multicolumn{4}{c|}{\textbf{Functional \CAS{}}} \\ \cline{3-10} 
\multicolumn{2}{|l|}{} & \textit{\textbf{Precision}} & \textit{\textbf{Recall}} & \textit{\textbf{F1}} & \textit{\textbf{\begin{tabular}[c]{@{}c@{}}Avg.\\ Count\end{tabular}}} & \textit{\textbf{Precision}} & \textit{\textbf{Recall}} & \textit{\textbf{F1}} & \textit{\textbf{\begin{tabular}[c]{@{}c@{}}Avg.\\ Count\end{tabular}}} \\ \hline\hline
\multicolumn{1}{|l|}{\multirow{7}{*}{\textbf{Reasoning LLM}}} & \textbf{Gemini-2.5-Pro} & 16.65 & 23.18 & 19.38 & 1.32 & \textbf{19.38} & 40.72 & \textbf{26.26} & 0.65 \\
\multicolumn{1}{|l|}{} & \textbf{Gemini-2.5-Flash} & \textbf{16.88} & 13.91 & 15.25 & 0.78 & 17.92 & 24.42 & 20.67 & 0.41 \\
\multicolumn{1}{|l|}{} & \textbf{DeepSeek-R1} & 14.61 & 25.5 & 18.58 & 1.66 & 15.55 & 50.62 & 23.79 & 1.06 \\
\multicolumn{1}{|l|}{} & \textbf{GPT-o3} & 14.05 & 25.58 & 18.13 & 1.73 & 14.41 & 50.78 & 22.45 & 1.13 \\
\multicolumn{1}{|l|}{} & \textbf{GPT-5} & 14.69 & 35.93 & \textbf{20.85} & 2.32 & 14.87 & \textbf{65.03} & 24.2 & 1.43 \\\cline{2-10}
\multicolumn{1}{|l|}{} & \textbf{Qwen-2.5-R1-32B} & 11.69 & 20.86 & 14.98 & 1.69 & 12.65 & 38.19 & 19 & 0.93 \\
\multicolumn{1}{|l|}{} & \textbf{Qwen-2.5-R1-14B} & 13.21 & 20.13 & 15.95 & 1.45 & 15.01 & 40.94 & 21.96 & 0.87 \\
\multicolumn{1}{|l|}{} & \textbf{Qwen-2.5-R1-7B} & 6.83 & 8.33 & 7.51 & 1.16 & 6.8 & 11.18 & 8.46 & 0.5 \\ \hline\hline
\multicolumn{1}{|l|}{\multirow{8}{*}{\textbf{Standard LLM}}} & \textbf{Claude-4-Opus} & 14.94 & 19.68 & 16.99 & 1.25 & 16.02 & 37.54 & 22.45 & 0.74 \\
\multicolumn{1}{|l|}{} & \textbf{Claude-4-Sonnet} & 13.84 & 20.76 & 16.61 & 1.42 & 14.81 & 39.45 & 21.54 & 0.87 \\
\multicolumn{1}{|l|}{} & \textbf{Claude-3.7-Sonnet} & 14.9 & 23.5 & 18.23 & 1.5 & 14.72 & 40.32 & 21.56 & 0.86 \\
\multicolumn{1}{|l|}{} & \textbf{GPT-4o} & 14.13 & \textbf{27.78} & 18.73 & \textbf{1.86} & 15.61 & 44.48 & 23.11 & 0.85 \\
\multicolumn{1}{|l|}{} & \textbf{DeepSeek-V3} & 15.49 & 20.17 & 17.52 & 1.22 & 16.15 & 33.56 & 21.81 & 0.61 \\ \cline{2-10} 
\multicolumn{1}{|l|}{} & \textbf{Qwen-2.5-32B} & 11.43 & 17.28 & 13.76 & 1.44 & 13.68 & 29.71 & 18.73 & 0.68 \\
\multicolumn{1}{|l|}{} & \textbf{Qwen-2.5-14B} & 8.67 & 9.38 & 9.01 & 1.03 & 13.65 & 17.68 & 15.41 & 0.4 \\
\multicolumn{1}{|l|}{} & \textbf{Qwen-2.5-7B} & 8.87 & 16.86 & 11.63 & 1.8 & 10 & 24.6 & 14.22 & 0.77 \\ \hline
\end{tabular}%
}
\end{table*}

\paragraph{Impact of LLMs and Reasoning Enhancement}

Following PR-Review's superior performance, we evaluated its efficacy on \swrbench{} with diverse LLMs, including variants with and without reasoning-enhancement training (Table~\ref{tab:llm_performance_pr_agent}).

Overall, LLM performance with PR-Review remains suboptimal for practical code review (highest $Overall\text{-}F1$ 19\%). Within this context, we first noted a significant divergence: the LLM performance ranking on \swrbench{} does not consistently mirror trends observed in other SE benchmarks (e.g., SWE-Bench~\cite{SWE-Bench} and LiveCodeBench~\cite{LiveCodeBench}). 
% Claude-4-Opus and Sonnet underperformed their predecessor Claude-3.7-Sonnet on \swrbench{}. 
This discrepancy underscores \swrbench{}'s value in capturing unique code review challenges and speculatively points to a potential misalignment between current LLM training/optimization and code review's specific demands, where overemphasis on other SE tasks might inadvertently reduce code review proficiency.

Turning to the impact of reasoning enhancement training, we found further nuances. Models like Gemini-2.5-Pro and DeepSeek-R1, known for their reasoning capabilities, generally performed better. The Qwen-2.5 series provided a clear illustration: its standard versions (without specific reasoning enhancement) yielded lower $F1$ scores (max 13.76\%). However, their reasoning-enhanced counterparts showed marked improvement. Specifically, Qwen-2.5-R1-14B achieved an $F1$ of 15.95\% (Qwen-2.5-R1-7B was an exception due to output formatting issues). This strongly suggests that reasoning-enhancement training is a crucial factor for improving LLM effectiveness in code review.

\mybox{Conclusion 7} % Renumbered Conclusion
{Current LLMs underperform for practical review on \swrbench{}; however, reasoning-enhanced models perform better, highlighting reasoning-enhanced as a crucial advancement path.}

\subsection*{RQ3: How can the performance of automated code review tools be improved?}

\begin{figure}[htbp] % 'h'ere, 't'op, 'b'ottom, 'p'age of floats. [!htbp] is more forceful.
    \centering % 中心对齐整个 figure 环境
    \begin{subfigure}[b]{0.48\linewidth} % 子图1，宽度为文本宽度的48%
        \centering
        \includegraphics[width=0.6\linewidth]{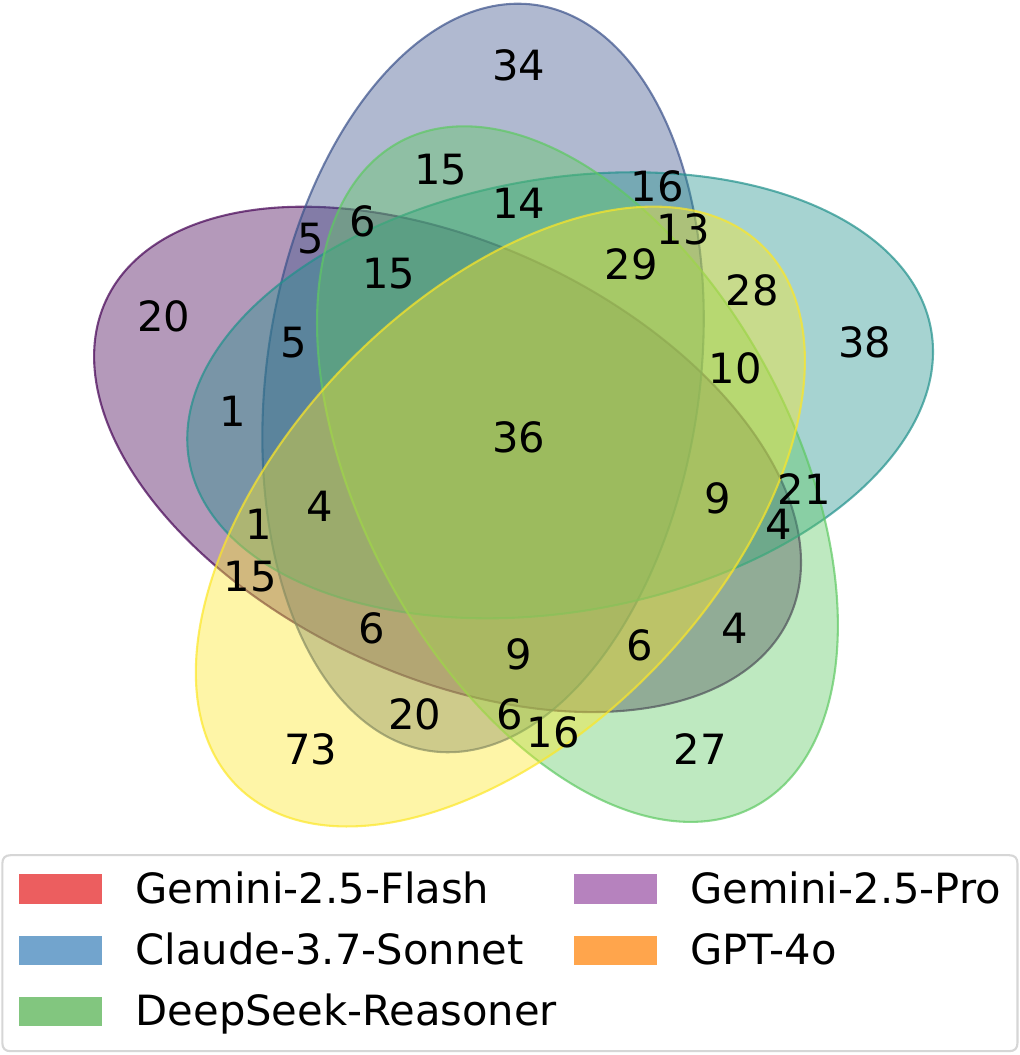} % 替换为你的第一个图片文件名
        \caption{Overlap across different models.}
        \label{fig:venn_diagram_model_overlap}
    \end{subfigure}
    \hfill % 在两个子图之间添加水平弹性空间，使它们分开
    \begin{subfigure}[b]{0.48\linewidth} % 子图2，宽度为文本宽度的48%
        \centering
        \includegraphics[width=0.6\linewidth]{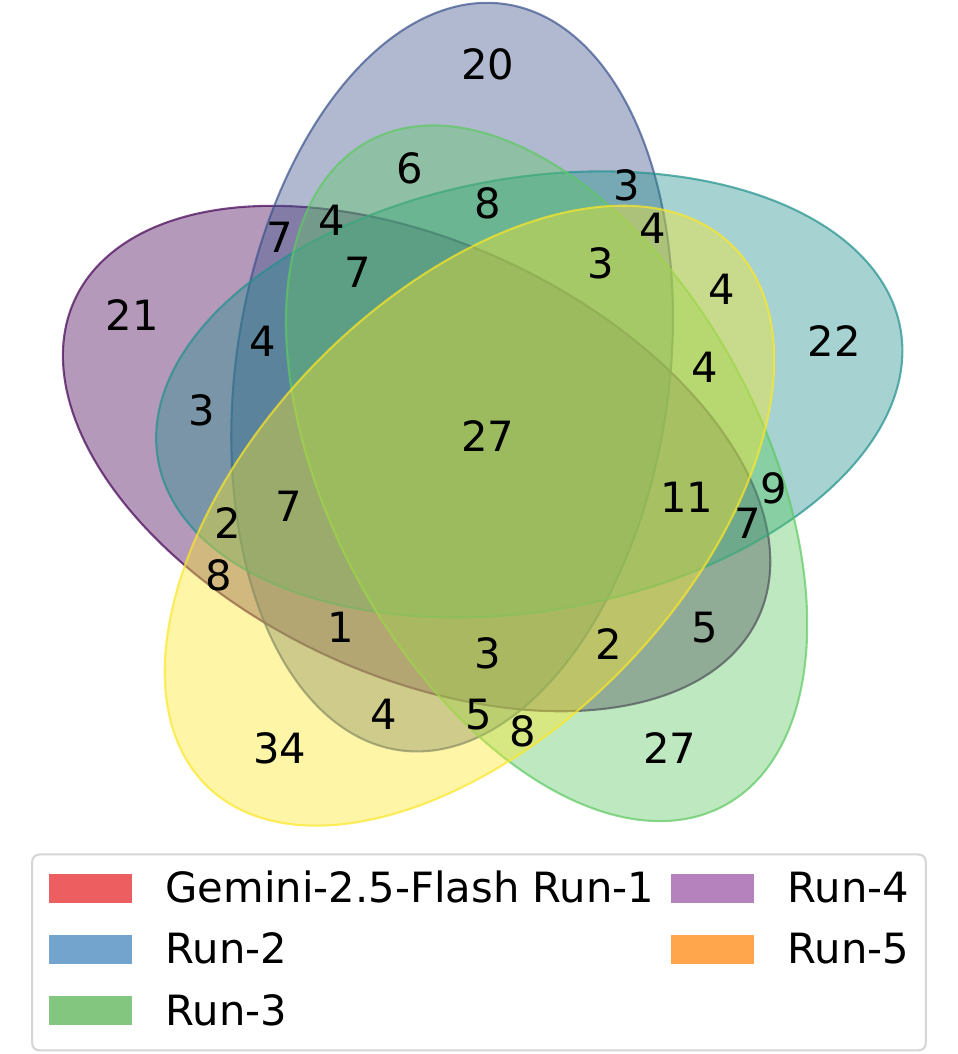} % 替换为你的第二个图片文件名
        \caption{Overlap across multiple runs of the same model.}
        \Description{}
        \label{fig:venn_diagram_run_overlap}
    \end{subfigure}
    \caption{Venn diagrams illustrating the overlap of identified \cas{}.} %(a) shows overlap between different models, and (b) shows overlap between multiple runs of the same model. % 整个大图的标题
    \label{fig:venn_diagrams_combined} % 整个大图的标签（可选，如果需要引用整个组合图）
\end{figure}

Conclusion 7 indicates reasoning-enhanced LLMs perform better, the overall practical utility of current ACR approaches remains limited. This prompted a closer examination of the nature of LLM-generated reviews, particularly their reliability and consistency. Specifically, we questioned whether current LLMs provide stable and consistent feedback across different invocations or when compared to other models.

To investigate this, we conducted an analysis into the overlap of specific ground truth \cas{} identified by different LLMs and by multiple runs of the same LLM using Venn diagrams (Figures~\ref{fig:venn_diagram_model_overlap} and~\ref{fig:venn_diagram_run_overlap}). This analysis revealed considerable variability and instability in the sets of successfully identified defects across these models and runs. Notably, for different LLMs only 36 successfully identified \cas{} overlapped (Figure~\ref{fig:venn_diagram_model_overlap}),  for the same LLM over five independent runs, only 27 successfully identified \cas{} overlapped (Figure~\ref{fig:venn_diagram_run_overlap}). This low consistency suggests unstable detection for most \cas{}, implying significant randomness in LLM identification. While this variability might partly stem from the subjective nature of many evolutionary changes, it strongly indicates that LLMs' current grasp of code review nuances may be superficial, with detections sometimes attributable to stochastic factors rather than deep comprehension.

\add{
\begin{figure}[htbp]
    \centering
    \includegraphics[width=0.75\linewidth]{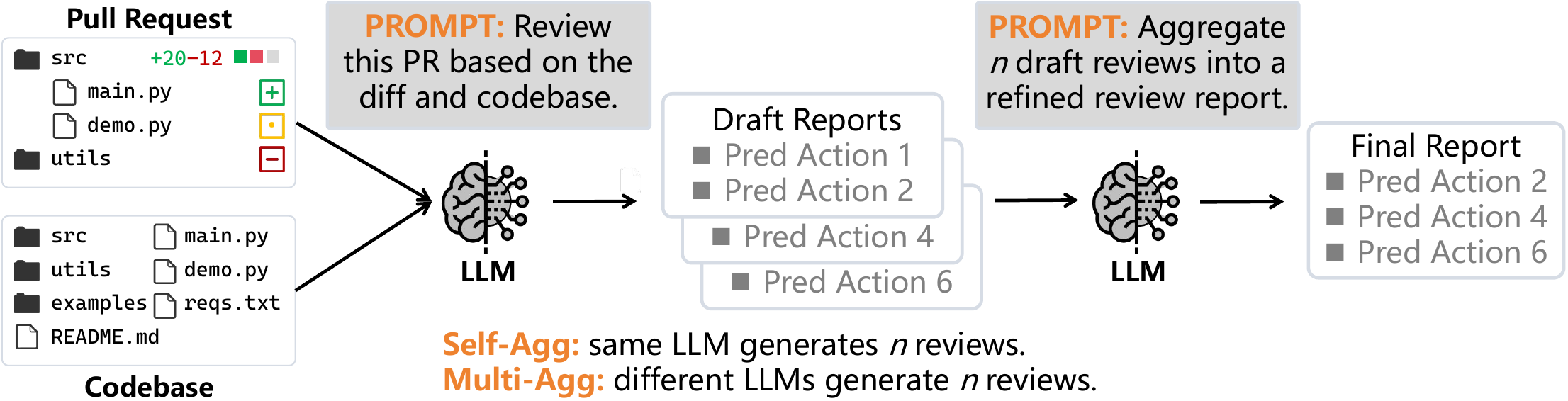}
    \caption{\note{BQ4: }\add{The overall workflow of the proposed Multi-Review approach.}}
    \label{fig:multi_review}
\end{figure}
}

This observed variability and the implied limitations of single-pass reviews directly motivated our exploration into whether integrating diverse review outputs could yield a more comprehensive and reliable final report. To this end, building upon PR-Review, we propose an enhanced approach named Multi-Review (Figure~\ref{fig:multi_review}). \note{BQ4: }\add{The core idea is to execute PR-Review (or another ACR tool) multiple times on the same code change to generate $n$ independent review reports, which are then aggregated into a final review report using an additional LLM call. Specifically, all $n$ reports are concatenated into a single input, and an LLM is prompted with a meta-instruction to synthesize them into a single, superior review. }
%The prompt instructs the LLM to: (1) identify issues consistently raised across multiple reports as high-confidence findings, (2) filter out issues that appear in only one report and lack sufficient justification, and (3) deduplicate redundant comments while preserving the most detailed and actionable version of each unique issue. The full aggregation prompt is provided in our replication package~\cite{swrbench}.
To validate Multi-Review's efficacy, we conducted experiments on \swrbench{} with two aggregation strategies: 1) \textbf{Self-Agg}, which employs the same LLM to both generate and aggregate multiple independent review reports, testing whether a single model can overcome its own stochasticity by consolidating multiple internal opinions; and 2) \textbf{Multi-Agg} (Cross-Model Aggregation), which aggregates reports generated by multiple different LLMs, leveraging the complementary strengths of different models to produce a more complete and accurate review.

% --- 合并后的图片代码 ---
\begin{figure*}[htbp] % 如果是双栏模板建议用 figure* 跨栏显示；如果是单栏改为 figure
    \centering
    % 子图 (a)
    \begin{subfigure}[b]{0.58\textwidth}
        \centering
        \includegraphics[width=\textwidth]{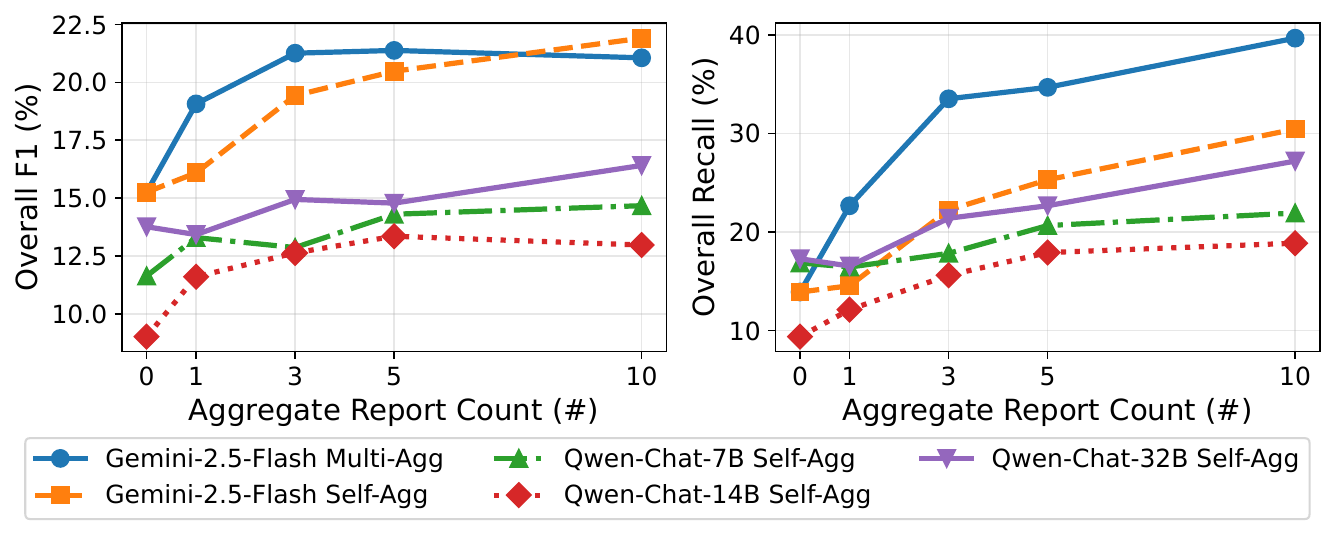}
        \caption{Performance of Multi-Review for varying numbers of aggregated reports ($n$).}
        \label{fig:mpr_agent_performance}
    \end{subfigure}
    \hfill
    % 子图 (b)
    \begin{subfigure}[b]{0.40\textwidth}
        \centering
        \includegraphics[width=\textwidth]{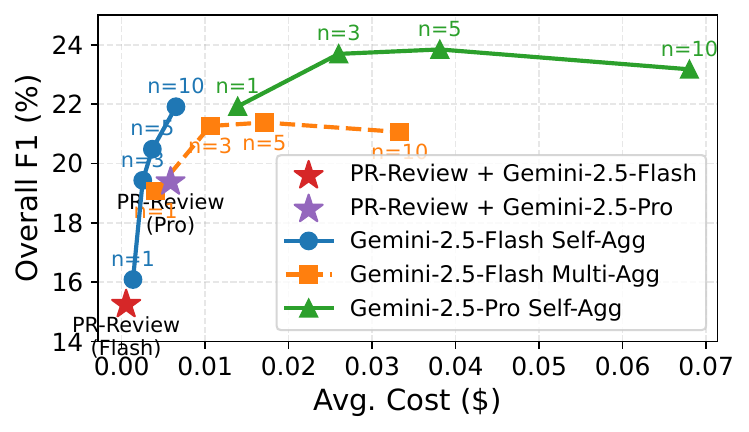}
        \caption{\note{MV3, AQ3, BQ4, CQ1, CQ2: }\add{Cost-benefit analysis across different models and aggregation sizes ($n$).}}
        \label{fig:rq3_efficiency}
    \end{subfigure}
    % 总标题
    \caption{Evaluation of the Multi-Review strategy. (a) shows the performance impact of aggregating varying numbers of reports, while (b) presents the trade-off between Overall F1 and average API cost per PR (in US dollars).}
    \label{fig:multi_review_combined}
\end{figure*}

% \begin{figure}[htbp] % 使用 table 环境，这样标题会是 "Table X"
%     \centering % 使图片居中
%     \includegraphics[width=0.60\linewidth]{figures/f1_score_combined.pdf} 
%     \caption{Performance of Multi-Review for varying numbers of aggregated reports ($n$).}
%     \Description{}
%     \label{fig:mpr_agent_performance} % 标签也保持一致
% \end{figure}

We selected Gemini-2.5-Flash as a representative LLM to investigate the impact of aggregating varying numbers of review reports ($n \in \{0, 1, 3, 5, 10\}$, where $n=0$ represents the baseline PR-Review without aggregation). The results, presented in Figure~\ref{fig:mpr_agent_performance}, demonstrate a significant performance uplift with Multi-Review. Regardless of whether aggregating results from different models or multiple runs of a single model, the $F1$ scores showed marked improvement. Specifically, Gemini-2.5-Flash with Self-Agg ($n=10$) achieved an $F1$ of 21.91\% (a 43.67\% increase) and a $Recall$ of 30.44\% (a 118.83\% increase). This larger gain in $Recall$ indicates Multi-Review's effectiveness in identifying more true defects. However, to further boost the $F1$ score, enhancing $Precision$ by reducing false positives remains a key priority.
To further assess the generalizability of Multi-Review, we experimented with smaller, open-source LLMs from the Qwen-2.5 series (7B, 14B, 32B), investigating whether these models could achieve significant self-enhancement via Multi-Review in resource-constrained scenarios. Specifically, Multi-Review consistently boosted $F1$ scores for the Qwen series. Notably, Qwen-Chat-7B Self-Agg ($n=10$) improved its $F1$ by 26.13\% (to 14.67\%), and Qwen-Chat-32B Self-Agg ($n=10$) enhanced its $F1$ by 19.25\% (to 16.41\%), allowing their aggregated results to approach the baseline PR-Review performance of some larger commercial LLMs.

% \begin{figure}[htbp]
% \centering
% % 请将 'efficiency_chart.png' 替换为您实际保存的图片文件名
% \includegraphics[width=0.5\linewidth]{figures/efficiency_analysis.pdf}
% \caption{\note{AQ3, BQ4, CQ1, CQ2: }\add{Cost-benefit analysis of Multi-Review. The plot shows the trade-off between Overall F1 and average API cost per PR (in US dollars) across different models and aggregation sizes ($n$).}}
% \label{fig:rq3_efficiency}
% \end{figure}

\note{MV3, AQ3, BQ4, CQ1, CQ2: }\add{Furthermore, Figure~\ref{fig:rq3_efficiency} presents a cost-benefit analysis of the Multi-Review, including results with the larger Gemini-2.5-Pro model. A key finding is that Gemini-2.5-Flash Self-Agg ($n$=5) achieves an F1 of 20.48\%, surpassing the single-pass Gemini-2.5-Pro baseline (F1: 19.38\%) at a lower API cost (\$3.68E-03 vs.\ \$5.86E-03). This demonstrates that multiple runs of a smaller, cheaper model can be more cost-effective than a single run of a larger, more expensive one. Meanwhile, Gemini-2.5-Pro Self-Agg ($n$=5) achieves the highest F1 of 23.84\% (a 23.01\% improvement over its single-pass baseline), confirming that the aggregation strategy benefits models across the entire performance spectrum. However, performance gains exhibit diminishing returns beyond $n$=5, while costs continue to scale linearly, suggesting that $n$=5 represents a practical sweet spot. Regarding time efficiency, generating initial drafts takes 38.75 seconds per PR. This step is fully parallelizable, meaning its duration remains constant regardless of $n$. The final aggregation step requires minimal additional time, taking only 4.67, 8.76, 12.80, and 24.35 seconds for $n$=1, 3, 5, and 10 reports respectively. This low time overhead confirms the approach is practical for real-world deployment.
}

\mybox{Conclusion 8}
{\add{Multi-Review improves code review performance across difference model sizes by aggregating multiple reports. Notably, aggregating several runs of a smaller model can cost-effectively match or exceed the accuracy of a single run from a larger model. This highlights the substantial potential of improving current tools through systematic output integration.}}

\section{Discussion and Implications}
\label{sec:discussion}

\note{AQ7, CQ4: }\add{The findings from our evaluation on \swrbench{} provide concrete insights into automated code review. We summarize these implications as follow:}

\noindent\textbf{Objective Evaluation through Fact Matching:}
\add{As demonstrated in Conclusion 1, traditional metrics like text similarity fail to reflect true semantic quality. Furthermore, relying on subjective large language model scoring to evaluate generated reviews shows poor agreement with human expert preferences. Researchers should therefore evaluate tools using objective fact matching against predefined ground truth to verify semantic issues reliably.}

\noindent\textbf{From Code Generation to Code Critique.}
\add{Code review is fundamentally different from code generation. Generation relies on pattern completion, whereas review demands logical deduction, counterfactual reasoning, and cross-file dependency analysis. This distinction explains why models optimized primarily for code generation underperform on \swrbench{}, and why reasoning-enhanced models consistently achieve better results (Conclusion~7). Future work should move beyond prompt engineering and explore dedicated training objectives for code critique, such as reinforcement learning with process reward models that evaluate the logical rigor of identified issues rather than only the correctness of generated code.}

\noindent\textbf{Mitigating False Positives in Practice.}
\add{The high false positive rates of current tools (Conclusion~2) severely limit their practical adoption by overwhelming developers with invalid alerts. Since most false positives arise from insufficient contextual understanding and ignorance of project-specific conventions (Table~\ref{tab:fp_categories}), future systems must integrate historical repository data to learn implicit coding norms. Furthermore, tool designers should decouple feedback based on severity. Because models detect functional changes more reliably than evolutionary ones (Conclusion~5), presenting functional defects as mandatory requirements and evolutionary suggestions as optional recommendations will significantly enhance the practical value of automated reviews.}

\noindent\textbf{Enhancing Review Performance through Report Aggregation:}
\add{Single passes of large language models exhibit significant randomness and miss different defects across runs. Aggregating multiple independent review reports into a single final output effectively mitigates this stochasticity and significantly improves overall defect detection. Conclusion 8 demonstrates that this aggregation strategy also allows multiple runs of smaller models to match or exceed the performance of a single run from a larger and more expensive model.}

\section{Threats to Validity}
\label{sec:threats}
\noindent\textbf{Internal Validity.} Threats to internal validity mainly arise from potential bugs in our implementation and the accuracy of manual verification. To mitigate these threats, we have conducted a detailed review of the code, and our code has been made publicly available at~\cite{swrbench} for independent verification.

% Threats to internal validity mainly arise from potential bugs in our implementation and the accuracy of manual verification. To mitigate these threats, we have conducted a detailed review of the code, and our code has been made publicly available at~\cite{swrbench} for independent verification.

\noindent\textbf{External Validity.} Threats to external validity mainly arise from the LLMs, ACR tools, and the quality of the selected projects. To mitigate these threats, we chose representative tools based on an extensive literature review and selected 12 popular, well-maintained open-source GitHub projects.

\noindent\textbf{Construct Validity.} The primary threat to construct validity centers on the accuracy and reliability of our objective LLM-based evaluation method. We mitigated this through human validation, which confirmed high consistency between our method's results and human expert evaluations.

\section{Conclusions}
\label{sec:conclusions}
This paper addressed critical limitations in existing automated code review benchmarks by introducing \swrbench{}, a novel benchmark featuring 1,000 real-world pull requests with full project context and an objective LLM-based evaluation. Our study on \swrbench{} revealed that while current LLM-based ACR systems generally underperform, they show better aptitude for functional error detection. To improve performance, we proposed and validated a multi-review aggregation strategy that significantly boosts $F1$ scores. \swrbench{}, along with our findings and proposed enhancement, provides a more realistic platform and valuable insights for advancing practical ACR research and development.

% \newpage{}
\section*{Data Availability}
The replication package for our study, containing the necessary source code and scripts to reproduce our experiments, is available at the anonymous repository~\cite{swrbench}.
% The experimental data supporting the findings of this paper has been submitted as supplementary material to the HotCRP review system. Due to the substantial size of the complete dataset, we have provided a representative subset of the data for review purposes. We are committed to making the full source code and the complete dataset publicly available in a permanent repository upon acceptance of this paper.

\bibliographystyle{ACM-Reference-Format}
\bibliography{ref}

\end{document}